\documentclass[10pt, pre, aps, twocolumn,showpacs,nofootinbib,longbibliography]{revtex4-1}
\usepackage{url,ulem}
\usepackage{amsmath,amssymb}
\usepackage[dvipdfmx]{graphicx,color}

\newcommand{\R}{\mathbb{R}}

\newcommand{\mM}{\boldsymbol{\rm M}}
\newcommand{\mK}{\boldsymbol{\rm K}}
\newcommand{\mL}{\boldsymbol{\rm L}}
\newcommand{\mS}{\boldsymbol{\rm S}}
\newcommand{\mA}{\boldsymbol{\rm A}}
\newcommand{\mB}{\boldsymbol{\rm B}}

\newcommand{\mO}{\boldsymbol{\rm O}}
\newcommand{\mP}{\boldsymbol{\rm P}}
\newcommand{\mX}{\boldsymbol{\rm X}}
\newcommand{\mLambda}{\boldsymbol{\rm \Lambda}}

\newcommand{\mE}{\boldsymbol{\rm E}}
\newcommand{\mW}{\boldsymbol{\rm W}}
\newcommand{\mN}{\boldsymbol{\rm N}}
\newcommand{\mQ}{\boldsymbol{\rm Q}}
\newcommand{\mJ}{\boldsymbol{\rm J}}
\newcommand{\mY}{\boldsymbol{\rm Y}}
\newcommand{\mZ}{\boldsymbol{\rm Z}}

\newcommand{\br}{\boldsymbol{r}}
\newcommand{\bl}{\boldsymbol{l}}
\newcommand{\btheta}{\boldsymbol{\theta}}
\newcommand{\bphi}{\boldsymbol{\phi}}
\newcommand{\bq}{\boldsymbol{q}}
\newcommand{\bp}{\boldsymbol{p}}
\newcommand{\be}{\boldsymbol{e}}
\newcommand{\bv}{\boldsymbol{v}}
\newcommand{\bx}{\boldsymbol{x}}

\newcommand{\by}{\boldsymbol{y}}

\newcommand{\bG}{\boldsymbol{G}}

\newcommand{\bs}{\boldsymbol{s}}
\newcommand{\bc}{\boldsymbol{c}}
\newcommand{\bd}{\boldsymbol{d}}

\newcommand{\bzero}{\boldsymbol{0}}

\newcommand{\bPhi}{\boldsymbol{\Phi}}
\newcommand{\norm}[1]{||#1||}
\newcommand{\dfracd}[2]{\dfrac{{\rm d} #1}{{\rm d} #2}}
\newcommand{\dfracp}[2]{\dfrac{\partial #1}{\partial #2}}
\newcommand{\dfracpp}[3]{\dfrac{\partial^{2} #1}{\partial #2\partial #3}}
\newcommand{\ave}[1]{\left\langle #1 \right\rangle}
\newcommand{\Tr}{{\rm Tr}}
\newcommand{\wt}{\widetilde}

\newcommand{\Vspring}{V_{\rm spring}}
\newcommand{\Vbend}{U_{\rm bend}}
\newcommand{\MatN}{{\rm Mat}(2N-2)}
\newcommand{\DiagN}{{\rm Diag}(2N-2)}

\begin{document}

\title{Mode selectivity of dynamically induced conformation in many-body chain-like bead-spring model}
\author{Yoshiyuki Y. Yamaguchi}
\email{yyama@amp.i.kyoto-u.ac.jp}
\affiliation{Department of Applied Mathematics and Physics, Graduate School of Informatics, Kyoto University, Kyoto 606-8501, Japan}

\begin{abstract}

  We consider conformation of a chain consisting of beads connected by stiff springs,
  where the conformation is determined by the bending angles
  between the consecutive two springs.
  A conformation is stabilized or destabilized not only by a given bending potential
  but also the fast spring motion,
  and stabilization by the spring motion depends on their excited normal modes.
  This stabilization mechanism has been named the dynamically induced conformation
  in a previous work on a three-body system.
  We extend analyses of the dynamically induced conformation
  in many-body chain-like bead-spring systems.
  The normal modes of the springs depend on the conformation,
  and the simple rule of the dynamical stabilization is
  that the lowest eigenfrequency mode contributes to the stabilization
  of the conformation.
  The high the eigenfrequency is, the more the destabilization emerges.
  We verify theoretical predictions by performing numerical simulations.
\end{abstract}
\maketitle

\section{Introduction}
\label{sec:Introduction}

Conformation has a deep relationship to function
as found in isomerization.
Maintenance of a conformation requires stability,
and the stability is usually associated
with landscape of the given potential energy function.
We however underline that dynamics also contributes to the stability.

A typical example of the dynamical stabilization is the Kapitza pendulum,
which is the inverted pendulum under the uniform gravity
\cite{stephenson-08,kapitza-51}.
The inverted pendulum is stabilized by applying fast vertical oscillation of the pivot.
This highly unintuitive stabilization is applied in a wide variety of fields
due to importance of the mechanism
\cite{bukov-dalessio-polkovnikov-15,grifoni-hanggi-96,wickenbrock-etal-12,chizhevsky-smeu-giacomelli-03,chizhevsky-14,cubero-etal-06,bordet-morfu-13,weinberg-14,uzuntarla-etal-15,buchanan-jameson-oedjoe-62,baird-63,jameson-66,apffel-20,bellman-mentsman-meerkov-86,shapiro-zinn-97,borromeo-marchesoni-07,richards-etal-18}.

In the Kapitza pendulum the stabilization is realized
by adding fast motion externally.
We stress that the fast oscillation is not necessarily external,
and an internal fast oscillation can also stabilize a conformation
in an autonomous system.
Indeed, in a chain-like bead-spring model consisting of beads connected by stiff springs,
the straight conformation is stabilized by fast spring motion,
whereas the potential energy function contains only the spring energy
and does not depend on the bending angles
\cite{yanagita-konishi-jp}.

The above dynamical stabilization in a bead-spring model
is firstly observed in numerical simulations,
and then theoretically analyzed in the three-body model \cite{yamaguchi-etal-22}
with the aids of the multiple-scale analysis \cite{bender-orszag-99}
and the averaging method
\cite{krylov-bogoliubov-34,krylov-bogoliubov-47,guckenheimer-holmes-83}.
A surprising result is that the stability of a conformation depends
on the excited normal modes of the springs.
Suppose that the system has the equal masses and the identical springs.
The straight (the fully bent) conformation is stabilized (destabilized)
by the in-phase mode and is destabilized (stabilized) by the antiphase mode.
Here, the in-phase (the antiphase) mode is a normal mode of the springs,
and is defined as the two springs expand and contract simultaneously (alternatively).
The stabilization is obtained from dynamics of the springs,
and the obtained conformation is called the dynamically induced conformation (DIC).

The previous analysis is performed in the three-body model,
and several natural questions are induced in many-body systems:
Is DIC ubiquitous?
How does the stability of a conformation depend on the excited normal modes?
Is there a simple rule in the dependence?
The aim of this paper is to answer these questions
in chain-like bead-spring models.

The present study has another importance in the context
of conformational isomerization in flexible molecules.
It is experimentally observed in $N$-acetyl-tryptophan methyl amide
that population of isomers is modified by exciting vibration in a bond,
and the modified population depends on the excited bond
\cite{dian-longarte-zweier-02},
whereas the Rice-Ramsperger-Kassel-Marcus theory 
\cite{RRKM1,RRKM2,RRKM3}
states that the destination is determined statistically.
This mode selectivity may have a deep connection with
the mode dependence of DIC.

This paper is organized as follows.
The chain-like bead-spring model is introduced in Sec.~\ref{sec:Model}
with the equations of motion.
We extract the equations of motion for the slow bending motion in Sec.~\ref{sec:EOM}.
Assuming absence of the bending potential for observing the simplest case,
we theoretically exhibit the excited mode dependence of stability
with concentrating on one-dimensional conformations,
whose bending angles are $0$ or $\pi$.
The theoretical predictions are examined through numerical simulations
in Sec.~\ref{sec:Numerics}
with applying the Lennard-Jones potential as the bending potential.
Finally, Sec.~\ref{sec:Summary} is devoted
to summary and discussions.

\section{Model}
\label{sec:Model}

\begin{figure}[h]
  \centering
  \includegraphics[width=7cm]{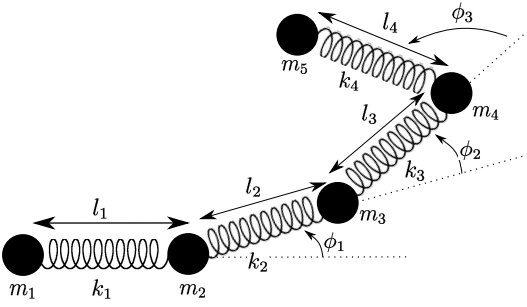}
  \caption{The bead-spring model on $\R^{2}$.
    This diagram shows an example of $N=5$ (five beads connected by four springs).}
  \label{fig:model}
\end{figure}

We consider the $N$-body chain-like bead-spring model on $\R^{2}$.
The model consists of $N$ beads connected by $N-1$ springs.
The $i$th bead is characterized by the mass $m_{i}\in (0,\infty)$,
the position $\br_{i}\in\R^{2}$, and the velocity 
$\dot{\br}_{i}={\rm d}\br_{i}/{\rm d}t \in\R^{2}$,
where $t\in\R$ is the time and $\br_{i}$ are column vectors.
See Fig.~\ref{fig:model} for a schematic diagram of the system.

\subsection{Lagrangian in Cartesian coordinates}

The Lagrangian of the system is
\begin{equation}
  \label{eq:L-in-r}
  L(\br,\dot{\br}) = K(\dot{\br}) - V(\br)
\end{equation}
 in the Cartesian coordinates, where
\begin{equation}
  \br =
  \begin{pmatrix}
    \br_{1} \\
    \vdots \\
    \br_{N}
  \end{pmatrix},
  \quad
  \dot{\br} =
  \begin{pmatrix}
    \dot{\br}_{1} \\
    \vdots \\
    \dot{\br}_{N} \\
  \end{pmatrix}
  \quad
  \in \R^{2N}.
\end{equation}
The kinetic energy function $K(\dot{\br})$ is
\begin{equation}
  \label{eq:K-in-r}
  K(\dot{\br}) = \dfrac{1}{2} \sum_{i=1}^{N} m_{i} \norm{\dot{\br}_{i}}^{2}.
\end{equation}
The potential energy function $V(\br)$ consists of
the spring potential $\Vspring$ and the bending potential $\Vbend$ as
\begin{equation}
  \label{eq:V-in-r}
  V(\br) = \Vspring(\br) + \Vbend(\br),
\end{equation}
where we assume that $\Vspring$ depends on only the lengths of the springs.
When we consider a molecule, the spring potential represents stronger bonds,
and the bending potential corresponds to weaker bonds.
A chain-like model is expressed by the nearest neighbor interactions
in the spring potential as
\begin{equation}
  \label{eq:Vspring}
  \Vspring(\br) = \sum_{i=1}^{N-1} V_{i}(\norm{\br_{i+1}-\br_{i}}).
\end{equation}

\subsection{Lagrangian in internal coordinates}

The system described by the Lagrangian \eqref{eq:L-in-r}
has the two-dimensional translational symmetry.
To reduce the two degrees of freedom,
we introduce $2N-2$ internal coordinates $\by\in\R^{2N-2}$ as
\begin{equation}
  \by =
  \begin{pmatrix}
    \bl \\
    \bphi \\
  \end{pmatrix}
  \in\R^{2N-2},
  \quad
  \left\{
    \begin{array}{l}
      \bl = (l_{1},\cdots,l_{N-1})^{\rm T} \in\R^{N-1} \\
      \bphi = (\phi_{1},\cdots,\phi_{N-1})^{\rm T} \in\R^{N-1} \\
    \end{array}
  \right.
\end{equation}
where the superscript T represents transposition.
The internal coordinates play an important role to extract
coupling between bending motion and spring motion
(see Ref.~\cite{yanao-etal-07} for instance).

The vector $\bl$ contains the lengths of the springs as
\begin{equation}
  l_{i} = \norm{\br_{i+1}-\br_{i}},
  \quad
  (i=1,\cdots,N-1).
\end{equation}
The Euclidean norm is defined by $\norm{\bx}=\sqrt{x^{2}+y^{2}}$ for $\bx=(x,y)\in\R^{2}$.

The angle $\phi_{i}~(i=1,\cdots,N-2)$ represents the bending angle
between the vectors $\br_{i+2}-\br_{i+1}$ and $\br_{i+1}-\br_{i}$,
and satisfies
\begin{equation}
  \cos\phi_{i}
  = \dfrac{(\br_{i+2}-\br_{i+1})\cdot (\br_{i+1}-\br_{i})}
  {\norm{\br_{i+2}-\br_{i+1}}~\norm{\br_{i+1}-\br_{i}}},
  \quad
  (i=1,\cdots,N-2)
\end{equation}
where $\bx\cdot\by$ is the Euclidean inner product between $\bx$ and $\by$.
The last angle variable $\phi_{N-1}$ associates to the total angular momentum.
The system has the rotational symmetry and $\phi_{N-1}$ is a cyclic coordinate,
but we keep it for later convenience.

Rewriting the kinetic energy in the variables $\by$ and $\dot{\by}$,
and assuming that the total angular momentum is zero,
we have the Lagrangian
\begin{equation}
  \label{eq:L-in-y}
  L(\by,\dot{\by})
  = \dfrac{1}{2} \sum_{\alpha,\beta=1}^{2N-2} B^{\alpha\beta}(\by)
  \dot{y}_{\alpha} \dot{y}_{\beta}
  - V(\by).
\end{equation}
We used the same symbol $V(\by)$ for the potential energy function
to simplify the notation.
The function $B^{\alpha\beta}(\by)$ is the $(\alpha,\beta)$ element
of the matrix $\mB(\by)\in\MatN$.
Here ${\rm Mat}(n)$ represents the set of real square matrices of size $n$.
The explicit form of $\mB(\by)$ is given in Appendix \ref{sec:L-in-y}.
Note that an alphabetic index runs from $1$ to $N-1$,
and a Greek index runs from $1$ to $2N-2$.

\subsection{Euler-Lagrange equations}
From now on, we adopt the Einstein notation for the sum:
We take the sum over an index if it appears twice in a term.
The Euler-Lagrange equation for the variable $y_{\alpha}~(\alpha=1,\cdots,2N-2)$
is expressed as
\begin{equation}
  \label{eq:EL-in-y}
  B^{\alpha\beta}(\by) \ddot{y}_{\beta}
  + D_{\alpha}^{\beta\gamma}(\by) \dot{y}_{\beta} \dot{y}_{\gamma}
  + \dfracp{V}{y_{\alpha}}(\by) = 0
\end{equation}
where
\begin{equation}
  D_{\alpha}^{\beta\gamma}(\by) =
  \dfracp{B^{\alpha\beta}}{y_{\gamma}}(\by)
  - \dfrac{1}{2} \dfracp{B^{\beta\gamma}}{y_{\alpha}}(\by).
\end{equation}

\section{Theory}
\label{sec:EOM}

This section provides a general theory to derive
the equations of motion which describe the slow bending motion.
We introduce two assumptions in Sec.~\ref{sec:assumptions}
with a dimensionless small parameter $\epsilon ~(0<\epsilon\ll 1)$.
The small parameter induces expansions of the variables $\by$ and $t$,
where the expansion of $t$ follows the multiple-scale analysis.
The spring and bending potentials are also expanded
in Sec.~\ref{sec:expansion-potentials}.
The slow bending motion is captured in $O(\epsilon^{2})$
of the expanded Euler-Lagrange equations
as shown in Sec.~\ref{sec:expanded-EL}.
To eliminate the fast timescale included in $O(\epsilon^{2})$,
we perform the averaging over the fast timescale in Sec.~\ref{sec:averaging}.
The averaged equations are not closed due to autonomy of the system,
and we will overcome this difficulty by introducing a hypothesis
in Sec.~\ref{sec:Hypothesis}
and the energy conservation law in Sec.~\ref{sec:energy-conservation}.
Finally, we will obtain closed equations 
for the slow bending motion in Sec.~\ref{sec:final-result}

The theory can be simplified in the chain-like model
with the aid of the explicit form of the spring potential \eqref{eq:Vspring}.
However, we develop the theory as general as possible in this section.

\subsection{Assumptions and expansions of variables}
\label{sec:assumptions}

Let $\bl_{\ast}=(l_{1,\ast},\cdots,l_{N-1,\ast})^{\rm T}$
be the natural length vector of the springs,
namely $\bl_{\ast}$ solves
\begin{equation}
  \dfracp{\Vspring}{\bl}(\bl_{\ast}) = \bzero.
\end{equation}
We introduce the two assumptions with a small parameter $\epsilon$:
\begin{itemize}
\item[{\it (A1)}] The amplitudes of the springs are sufficiently small comparing
  with the natural lengths. The ratio is of $O(\epsilon)$.
\item[{\it (A2)}] Large bending motion is sufficiently slow than the spring motion.
  The ratio of the two timescales is of $O(\epsilon)$.
\end{itemize}
We express these assumptions by the expansions of $t, \bl$, and $\bphi$ as 
\begin{equation}
  \label{eq:expansion-t}
  \dfracd{}{t} = \dfracp{}{t_{0}} + \epsilon \dfracp{}{t_{1}}
\end{equation}
and
\begin{equation}
  \label{eq:expansion-l-phi}
  \begin{split}
    & \bl(t_{0},t_{1}) = \bl_{\ast} + \epsilon \bl^{(1)}(t_{0},t_{1}), \\
    & \bphi(t_{0},t_{1}) = \bphi^{(0)}(t_{1}) + \epsilon \bphi^{(1)}(t_{0},t_{1}), \\
  \end{split}
\end{equation}
which is summarized as
\begin{equation}
  \by(t_{0},t_{1}) = \by^{(0)}(t_{1}) + \epsilon \by^{(1)}(t_{0},t_{1}).
\end{equation}
The two timescales $t_{0}=t$ and $t_{1}=\epsilon t$
correspond to the fast spring motion and the slow bending motion, respectively.
The vector $\bl_{\ast}=(l_{1,\ast},\cdots,l_{N-1,\ast})^{\rm T}$ is constant.
We are interested in the large and slow motion of the bending angles
described by $\bphi^{(0)}(t_{1})$.

Two remarks are in order.
First, the leading order of the expanded equations of motion is of $O(\epsilon)$,
since the leading order of the velocities and the accelerations is of $O(\epsilon)$.
Second, the velocities $\dot{l}_{i}$ and $\dot{\phi}_{i}$
are of the same order of $O(\epsilon)$,
and the fastness of motion in the assumption {\it (A2)} connotes a short period.
Indeed, the distance of a normal mode orbit is of $O(\epsilon)$
and the period is of $O(\epsilon^{0})$,
while the distance of the large bending motion is of $O(\epsilon^{0})$
and the period is of $O(\epsilon^{-1})$.

\subsection{Expansions of the potential function}
\label{sec:expansion-potentials}
We also expand the potential functions $\Vspring$ and $\Vbend$
into power series of $\epsilon$.
The expansions of the two functions are performed in different ways,
as a result of Eq.~\eqref{eq:expansion-l-phi}.

The spring potential function $\Vspring$ is expanded around $\bl_{\ast}$ as
\begin{equation}
  \label{eq:V-expansion}
  \Vspring(\bl) = \Vspring(\bl_{\ast})
  + \dfrac{1}{2} (\bl-\bl_{\ast})^{\rm T} \mK_{l} (\bl-\bl_{\ast}) + O(\epsilon^{3}),
\end{equation}
where the $(i,j)$ element $K_{l}^{ij}$ of the matrix $\mK_{l}\in{\rm Mat}(N-1)$ is
\begin{equation}
  K_{l}^{ij} = \dfracpp{\Vspring}{l_{i}}{l_{j}}(\bl_{\ast}),
  \quad
  (i,j=1,\cdots,N-1)
\end{equation}
and $\mK_{l}$ is assumed to be positive definite.

The bending potential $\Vbend$ can be also expanded around
a stationary point $\bphi_{\ast}$ if it exists,
but this expansion is not useful since
$\norm{\bphi-\bphi_{\ast}}$ is not necessarily of $O(\epsilon)$.
We thus expand $\Vbend$ by its amplitude as
\begin{equation}
  \label{eq:U-expansion}
  \Vbend(\by) = \Vbend^{(0)}(\by) + \epsilon \Vbend^{(1)}(\by)
  + \epsilon^{2} \Vbend^{(2)}(\by) + \cdots.
\end{equation}
It is important to note that $\Vbend^{(0)}(\by)$ and $\Vbend^{(1)}(\by)$
solely depend on $\bl$ with the conditions
\begin{equation}
  \dfracp{\Vbend^{(0)}}{\bl}(\bl_{\ast}) = \dfracp{\Vbend^{(1)}}{\bl}(\bl_{\ast}) = \bzero
\end{equation}
for satisfying the assumptions
(see Appendix \ref{sec:ordering-bending-potential}).
Integrating them into the spring potential $\Vspring(\bl)$,
we may set $\Vbend^{(0)}(\by),\Vbend^{(1)}(\by)\equiv 0$.
The leading order of $\Vbend$ is hence of $O(\epsilon^{2})$,
and this ordering is consistent with weaker bonds derived from the bending potential.

\subsection{Expansion of the Euler-Lagrange equations}
\label{sec:expanded-EL}

Substituting Eqs.~\eqref{eq:expansion-t}, \eqref{eq:expansion-l-phi},
\eqref{eq:V-expansion}, and \eqref{eq:U-expansion}
into Eq.~\eqref{eq:EL-in-y}, we have the expanded equations of motion
order by order of $\epsilon$.
As remarked in the end of Sec.~\ref{sec:assumptions},
the nontrivial equations of motion start from $O(\epsilon)$.

\subsubsection{$O(\epsilon)$}

The equations of motion in $O(\epsilon)$ is linear and
governed by the springs as
\begin{equation}
  \label{eq:EOM-epsilon1}
  \dfracp{{}^{2}\by^{(1)}}{t_{0}^{2}} = - \mX(\by^{(0)}) \by^{(1)}
\end{equation}
where
\begin{equation}
  \mX(\by) = [ \mB(\by) ]^{-1} \mK \in \MatN
\end{equation}
and
\begin{equation}
  \mK =
  \begin{pmatrix}
    \mK_{l} & \mO \\
    \mO & \mO
  \end{pmatrix}
  \in \MatN.
\end{equation}
The symbol $\mO$ represents the zero matrix.
The variables $\bphi^{(0)}$ to observe do not appear in $O(\epsilon)$,
and we progress to the next order.

\subsubsection{$O(\epsilon^{2})$}
In $O(\epsilon^{2})$ the equations of motion are
\begin{equation}
  \label{eq:EOM-epsilon2}
  \begin{split}
    & B^{\alpha\beta}(\by^{(0)}) \left( \ddot{y}_{\beta} \right)^{(2)}
    + D_{\alpha}^{\beta\gamma}(y^{(0)}) 
    \left( \dot{y}_{\beta} \right)^{(1)} \left( \dot{y}_{\gamma} \right)^{(1)} \\
    &
    + \dfracp{B^{\alpha\beta}}{y_{\gamma}}(y^{(0)})
    \left( \ddot{y}_{\beta} \right)^{(1)} y_{\gamma}^{(1)}
    + \dfracp{\Vbend^{(2)}}{y_{\alpha}}(\by^{(0)})
    = 0, 
  \end{split}
\end{equation}
where
\begin{equation}
  \label{eq:derivative-order}
  \begin{split}
    & \left( \dot{\by} \right)^{(1)}
    = \dfracd{\by^{(0)}}{t_{1}} + \dfracp{\by^{(1)}}{t_{0}}, 
    \qquad
    \left( \ddot{\by} \right)^{(1)}
    = \dfracp{{}^{2}\by^{(1)}}{t_{0}^{2}}, \\
    & \left( \ddot{\by}\right)^{(2)}
    = \dfracd{{}^{2}\by^{(0)}}{t_{1}^{2}} + 2 \dfracpp{\by^{(1)}}{t_{0}}{t_{1}}. \\
  \end{split}
\end{equation}
The vector $(\dot{\by})^{(1)}$ is the first order part of $\dot{\by}$
and $(\dot{\by})^{(1)}\neq {\rm d}\by^{(1)}/{\rm d}t$.
The variables $\bphi^{(0)}$ appear in $(\dot{\by})^{(1)}$ and $(\ddot{\by})^{(2)}$.

\subsection{Averaging}
\label{sec:averaging}

The equations \eqref{eq:EOM-epsilon2} depend on the fast timescale $t_{0}$
through $\by^{(1)}$, and we eliminate it by taking the average.
The average in the timescale $t_{0}$ is defined by
\begin{equation}
  \ave{\varphi} = \lim_{T\to\infty} \dfrac{1}{T} \int_{0}^{T} \varphi(t_{0}) dt_{0}.
\end{equation}
The averaged equations are
\begin{equation}
  \label{eq:EOM-epsilon2-averaged}
  \begin{split}
    & B^{\alpha\beta}(\by^{(0)}) \dfracd{{}^{2}y_{\beta}^{(0)}}{t_{1}^{2}}
    + D_{\alpha}^{\beta\gamma}(\by^{(0)})
    \dfracd{y_{\beta}^{(0)}}{t_{1}} \dfracd{y_{\gamma}^{(0)}}{t_{1}} \\
    & + \dfracp{\Vbend^{(2)}}{y_{\alpha}}(\by^{(0)})
    = \mathcal{A}_{\alpha},
  \end{split}
\end{equation}
where the averaged term $\mathcal{A}_{\alpha}$ is
\begin{equation}
  \label{eq:mathcalA}
  \begin{split}
    \mathcal{A}_{\alpha}
    & = \dfrac{1}{2} \Tr \left[ \dfracp{\mB}{y_{\alpha}}(\by^{(0)})
      \mX(\by^{(0)}) \ave{ \by^{(1)} \by^{(1){\rm T}} }  \right]. \\
  \end{split}
\end{equation}
The symbol $\Tr$ represents the matrix trace.
We used the relation
\begin{equation}
  \label{eq:relation}
  \ave{ \dfracp{{\by}^{(1)}}{t_{0}} \left( \dfracp{\by^{(1)}}{t_{0}} \right)^{\rm T} }
  = \mX(\by^{(0)}) \ave{ \by^{(1)} \by^{(1){\rm T}} } 
\end{equation}
proven by performing the integration by parts.

The averaged term $\mathcal{A}_{\alpha}$ depends on the solution $\by^{(1)}$
to Eq.~\eqref{eq:EOM-epsilon1}, which is obtained
by diagonalizing the matrix $\mX(\by^{(0)})$ as
\begin{equation}
  \label{eq:diagonalization}
  \mX(\by^{(0)}) \mP(\by^{(0)}) = \mP(\by^{(0)}) \mLambda(\by^{(0)}).
\end{equation}
$\mP\in\MatN$ is a diagonalizing matrix, and the diagonal matrix
$\mLambda$ contains the eigenvalues of $\mX$:
\begin{equation}
  \mLambda(\by^{(0)}) = {\rm diag}(\lambda_{1},\cdots,\lambda_{N-1},0,\cdots,0)
  \in\DiagN,
\end{equation}
where the symbol ${\rm Diag}(n)$ represents the set of real diagonal matrices.
The $N-1$ zeroeigenvalues come from the fact
$\dim({\rm Ker}\mK)=N-1$ in general.
We assume that the corresponding $N-1$ amplitudes of $\by^{(1)}$ are zero.
Since the average of the square of a sinusoidal function is $1/2$, we have in general
\begin{equation}
  \label{eq:average-y1y1}
  \ave{\by^{(1)}\by^{(1){\rm T}}}
  = \dfrac{1}{2} \mP(\by^{(0)}) \mW(t_{1})^{2} \mP(\by^{(0)})^{\rm T},
\end{equation}
where
\begin{equation}
  \mW(t_{1}) = {\rm diag}(w_{1},\cdots,w_{N-1},0,\cdots,0) \in \DiagN.
\end{equation}
The matrix $\mW$ contains the $N-1$ nontrivial amplitudes $w_{i}~(i=1,\cdots,N-1)$,
which evolve in the slow timescale $t_{1}$ through coupling with $\bphi^{(0)}(t_{1})$,
while the initial phases of $\by^{(1)}$ are eliminated by the average.

Summarizing, the averaged term $\mathcal{A}_{\alpha}$ is modified
by Eqs.~\eqref{eq:diagonalization} and \eqref{eq:average-y1y1} to
\begin{equation}
  \label{eq:mathcalA-averaged}
  \begin{split}
    & \mathcal{A}_{\alpha}(\by^{(0)})
    = \dfrac{1}{4} \Tr \left[
      \mP^{\rm T} \dfracp{\mB}{y_{\alpha}} \mP \mLambda \mW(t_{1})^{2}
    \right], \\
    & (\alpha=1,\cdots,2N-2).
  \end{split}
\end{equation}
We can show that
\begin{equation}
  \label{eq:mathcalA-zeros}
  \mathcal{A}_{1} = \cdots = \mathcal{A}_{N-1} = 0.
\end{equation}
In other words, the averaged terms survive only in the equations for $\bphi^{(0)}$.
See Appendix \ref{sec:simplifications}.

\subsection{Hypothesis}
\label{sec:Hypothesis}

We stress that Eq.~\eqref{eq:EOM-epsilon2-averaged} is not closed,
because the averaged term \eqref{eq:mathcalA-averaged} depends on $\mW(t_{1})$,
which nontrivially evolves in the timescale $t_{1}$.
We have to eliminate the $N-1$ nontrivial amplitudes $w_{i}(t_{1})~(i=1,\cdots,N-1)$.
To this end, we introduce the hypothesis \cite{yamaguchi-etal-22}
inspired from the adiabatic invariance:
\begin{equation}
  \label{eq:Hypothesis}
  w_{i}(t_{1})^{2} = \nu_{i} w(t_{1})^{2}
  \quad
  (i=1,\cdots,N-1).
\end{equation}
This hypothesis supposes that the normal mode energy ratios are constant of time,
and reduces the number of unknown variables from $N-1$ to one,
although the averaged term $\mathcal{A}_{\alpha}$ depends on the constants
$\nu_{i}~(i=1,\cdots,N-1)$.
We arrange the constants in the matrix
\begin{equation}
  \label{eq:Hypothesis-Matrix}
  \mN = {\rm diag}(\nu_{1},\cdots,\nu_{N-1},0,\cdots,0) \in \DiagN,
\end{equation}
and the amplitude matrix $\mW$ is simplified to
\begin{equation}
  \mW(t_{1})^{2} = w(t_{1})^{2} \mN.
\end{equation}
The averaged term $\mathcal{A}_{\alpha}$ is then modified to
\begin{equation}
  \label{eq:mathcalA-hypothesis}
  \mathcal{A}_{\alpha}(\by^{(0)})
  = \dfrac{w(t_{1})^{2}}{4} \Tr \left[
    \mP^{\rm T} \dfracp{\mB}{y_{\alpha}} \mP \mLambda \mN
  \right].
\end{equation}

We remark that the numbering of the normal modes is important
in the application of the hypothesis,
since the eigenvalues of $\mX(\by^{(0)})$ depend on the bending angles $\bphi^{(0)}$.
In $N=3$, the two springs expand and contract simultaneously (alternatively)
in the in-phase mode (the antiphase mode),
which has the energy ratio $\nu_{\rm in}$ ($\nu_{\rm anti}$).
Adopting $\nu_{1}=\nu_{\rm in}$ and $\nu_{2}=\nu_{\rm anti}$,
the hypothesis \eqref{eq:Hypothesis} is approximately verified
for any value of $\bphi^{(0)}$ \cite{yamaguchi-etal-22}.

\begin{figure}[htb]
  \centering
  \includegraphics[width=8cm]{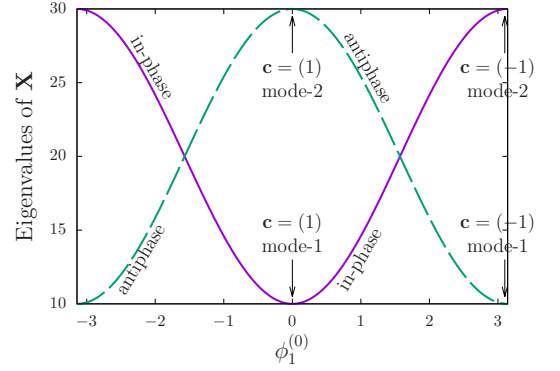}
  \caption{The two eigenvalues of $\mX(\by^{(0)})$,
    which depend on $\phi_{1}^{(0)}$ for $N=3$.
    The in-phase mode (purple solid line) and the antiphase mode (green dashed line).
    $m_{1}=m_{2}=m_{3}=m=1$. $\mK_{l}=k\mE$ with $k=10$,
    where $\mE$ is the unit matrix.
    The eigenvalues are $(k/m)(2\mp\cos\phi_{1}^{(0)})$.
    The modes are locally numbered in the ascending order of the eigenvalues.
    The conformation symbol $\bc$ is defined in Sec.~\ref{sec:1Dconf}:
    $\bc=(1)$ represents the straight conformation and
    $\bc=(-1)$ the fully bent conformation.
  }
  \label{fig:XEigenvaluesN3}
\end{figure}

However, the global numbering is not trivial in general,
because two eigenvalues of $\mX(\by^{(0)})$ may cross by varying $\bphi^{(0)}$
as shown in Fig.~\ref{fig:XEigenvaluesN3}.
Nevertheless, the hypothesis is approximately valid when the bending motion
is not large, since the system is close to the integrable system, Eq.~\eqref{eq:EOM-epsilon1}, and the mode numbers can be identified in a local region of $\bphi^{(0)}$.
Consequently, the hypothesis is useful to study
stationarity and stability of a conformation.
From now on, 
we locally number the modes in the ascending order of the eigenvalues
(see Fig.~\ref{fig:XEigenvaluesN3}) unless there otherwise stated.

\subsection{Energy conservation}
\label{sec:energy-conservation}

The last unknown variable $w(t_{1})$ is eliminated by the energy conservation.
Expanding energy $E$ as $E=\epsilon^{2}E^{(2)}+\cdots$,
the leading term is written as
\begin{equation}
  \label{eq:E2}
  \begin{split}
    E^{(2)}
    & = \dfrac{1}{2} \Tr \left[ \mB(\by^{(0)})
      \left( \dot{\by} \right) ^{(1)} \left( \dot{\by} \right)^{(1){\rm T}} \right]
    + \Vbend^{(2)}(\by^{(0)}) \\
    & + \dfrac{1}{2} \Tr \left[ \mK \by^{(1)} \by^{(1){\rm T}} \right].
  \end{split}
\end{equation}
Taking the average, we have
\begin{equation}
  \label{eq:aveE2}
  \begin{split}
    \ave{E^{(2)}}
    & = \dfrac{1}{2} \Tr \left[ \mB(\by^{(0)})
      \dfracd{\by^{(0)}}{t_{1}} \left( \dfracd{\by^{(0)}}{t_{1}} \right)^{\rm T} ~ \right]
    + \Vbend^{(2)}(\by^{(0)}) \\
    & + \dfrac{1}{2} \Tr \left[ \mP^{\rm T} \mK \mP \mW^{2} \right]  \\
  \end{split}
\end{equation}
by the relations \eqref{eq:relation} and \eqref{eq:average-y1y1}.
In the right-hand side of Eq.~\eqref{eq:aveE2},
the first line represents the bending energy,
and the second line the spring energy.

The hypothesis \eqref{eq:Hypothesis-Matrix} gives the equality
\begin{equation}
  \label{eq:wt1}
  \begin{split}
    w(t_{1})^{2} \Tr\left[ \mP^{\rm T} \mK \mP \mN \right]
    & = 2 \left[ E^{(2)} - \Vbend^{(2)}(\by^{(0)}) \right]  \\
    & - \Tr \left[ \mB(\by^{(0)})
      \dfracd{\by^{(0)}}{t_{1}} \left( \dfracd{\by^{(0)}}{t_{1}} \right)^{\rm T} ~ \right],
  \end{split}
\end{equation}
where we denoted $\ave{E^{(2)}}$ by $E^{(2)}$ for simplicity.
Substituting Eq.~\eqref{eq:wt1} into Eq.~\eqref{eq:mathcalA-hypothesis},
we have the averaged term $\mathcal{A}_{\alpha}$ as
\begin{equation}
  \label{eq:mathcalA-final}
  \begin{split}
    \mathcal{A}_{\alpha}(\by^{(0)})
    & = \left\{
      \dfrac{E^{(2)}-\Vbend^{(2)}(\by^{(0)})}{2}
    \right. \\
    & \left. - \dfrac{1}{4} \Tr \left[ \mB(\by^{(0)})
        \dfracd{\by^{(0)}}{t_{1}} \left( \dfracd{\by^{(0)}}{t_{1}} \right)^{\rm T} ~ \right]
    \right\} \mathcal{S}_{\alpha}(\by^{(0)})
  \end{split}
\end{equation}
where $\mathcal{S}_{\alpha}~(\alpha=1,\cdots,2N-2)$ are
\begin{equation}
  \mathcal{S}_{\alpha}(\by)
  = \dfrac{ \Tr \left[
      \mP(\by)^{\rm T} \dfracp{\mB}{y_{\alpha}}(\by) \mP(\by) \mLambda(\by) \mN \right] }
  {\Tr \left[ \mP(\by)^{\rm T} \mK \mP(\by) \mN \right]}.
\end{equation}
We can show that
\begin{equation}
  \label{eq:S1-SNm1-vanish}
  \mathcal{S}_{1} = \cdots = \mathcal{S}_{N-1} = 0.
\end{equation}
See Appendix \ref{sec:simplifications}.
We remark that in the inside traces of $\mathcal{S}_{\alpha}$
the size of the matrices can be reduced from $2N-2$ to $N-1$
as derived in Appendix \ref{sec:simplifications}.

We have eliminated the nontrivial unknown variables $w_{i}(t_{1})~(i=1,\cdots,N-1)$
from the averaged term \eqref{eq:mathcalA-final},
which depends on only the variables 
$\bphi^{(0)}(t_{1})$ and the constants $\bl_{\ast}$, $\mN$, and $E^{(2)}$.
Therefore, the equations of motion \eqref{eq:EOM-epsilon2-averaged}
for $\bphi^{(0)}$ is now closed,
and dynamics depends on the excited normal modes $\mN$
and energy $E^{(2)}$.

\subsection{Final result}
\label{sec:final-result}

Substituting Eq.~\eqref{eq:mathcalA-final} into
Eq.~\eqref{eq:EOM-epsilon2-averaged}, we have
\begin{widetext}
  \begin{equation}
  \begin{split}
    & B^{\alpha\beta}(\by^{(0)}) \dfracd{{}^{2}y_{\beta}^{(0)}}{t_{1}^{2}}
    + \left[ D_{\alpha}^{\beta\gamma}(\by^{(0)})
      + \dfrac{1}{4} B^{\beta\gamma}(\by^{(0)}) \mathcal{S}_{\alpha}(\by^{(0)}) \right]
    \dfracd{y_{\beta}^{(0)}}{t_{1}} \dfracd{y_{\gamma}^{(0)}}{t_{1}} 
    +  \dfracp{\Vbend^{(2)}}{y_{\alpha}}(\by^{(0)}) 
    - \dfrac{E^{(2)}-\Vbend^{(2)}(\by^{(0)})}{2}
    \mathcal{S}_{\alpha}(\by^{(0)}) = 0
  \end{split}
\end{equation}
\end{widetext}
for $\alpha=1,\cdots,2N-2$.
The final result for the bending variables $\bphi^{(0)}(t_{1})$ is
\begin{equation}
  \label{eq:EOM-slow}
  \dfracd{{}^{2}\phi_{i}^{(0)}}{t_{1}^{2}}
  + F_{i}^{jn}(\by^{(0)}) \dfracd{\phi_{j}^{(0)}}{t_{1}} \dfracd{\phi_{n}^{(0)}}{t_{1}}
  + G_{i}(\by^{(0)}) = 0
\end{equation}
for $i=1,\cdots,N-1$.
The functions $F_{i}^{jn}$ and $G_{i}$ are 
\begin{equation}
  \label{eq:Fijn}
  F_{i}^{jn}(\by)
  = \left[ B_{\phi\phi}^{-1} \right]^{is}
  \left(
    \dfracp{B_{\phi\phi}^{sj}}{\phi_{n}}
    - \dfrac{1}{2} \dfracp{B_{\phi\phi}^{jn}}{\phi_{s}}
    + \dfrac{1}{4} \mathcal{T}_{s} B_{\phi\phi}^{jn}
  \right)
\end{equation}
and
\begin{equation}
  \label{eq:Gi}
  G_{i}(\by)
  = \left[ B_{\phi\phi}^{-1} \right]^{is}
  \left(
    \dfracp{\Vbend^{(2)}}{\phi_{s}}
    - \dfrac{E^{(2)}-\Vbend^{(2)}}{2}
    \mathcal{T}_{s}
  \right).
\end{equation}
Here, we decomposed the matrix $\mB\in\MatN$
into four square submatrices of the size $N-1$ as
\begin{equation}
  \mB(\by) =
  \begin{pmatrix}
    \mB_{ll}(\bphi) & \mB_{l\phi}(\by) \\
    \mB_{\phi l}(\by) & \mB_{\phi\phi}(\by) \\
  \end{pmatrix}.
\end{equation}
See Eq.~\eqref{eq:B-submatrices} for explicit forms of the submatrices.
The functions
\begin{equation}
  \begin{split}
    & \mathcal{T}_{i}(\by) = \mathcal{S}_{i+N-1}(\by)
    = \dfrac{ \Tr \left[
        \mP^{\rm T} \dfracp{\mB}{\phi_{i}} \mP \mLambda \mN \right] }
    {\Tr \left[ \mP^{\rm T} \mK \mP \mN \right]} \\
    & (i=1,\cdots,N-1)
  \end{split}
\end{equation}
are introduced to renumber $\mathcal{S}_{\alpha}$
for avoiding zerocomponents shown in Eq.~\eqref{eq:S1-SNm1-vanish}.
It is worth noting that the spring potential $\Vspring$ is included
in the final equations of motion \eqref{eq:EOM-slow} up to the second order,
namely only through the matrix $\mK_{l}$.

\section{Dynamically induced stability}
\label{sec:stationarity-stability}

The equations \eqref{eq:EOM-slow} are rewritten as
\begin{equation}
  \label{eq:EOM-in-normalform}
  \dfracd{\phi_{i}^{(0)}}{t_{1}} = v_{i},
  \quad
  \dfracd{v_{i}}{t_{1}} = - 
  F_{i}^{jn}(\by^{(0)}) v_{j} v_{n} - G_{i}(\by^{(0)}),
\end{equation}
which describe dynamics on the $2N-2$ dimensional space of
\begin{equation}
  \bPhi =
  \begin{pmatrix}
    \bphi^{(0)} \\
    \bv
  \end{pmatrix}
  \in \R^{2N-2},
\end{equation}
where $\bv=(v_{1},\cdots,v_{N-1})^{\rm T}$.
It is clear that
\begin{equation}
  \bPhi_{\ast} =
  \begin{pmatrix}
    \bphi^{(0)}_{\ast} \\ \bv_{\ast}
  \end{pmatrix}
  : \text{stationary}
  \quad\Longleftrightarrow\quad
  \left\{
    \begin{array}{l}
      \bG(\by^{(0)}_{\ast}) = \bzero, \\
      \bv_{\ast} = \bzero.
    \end{array}
  \right.
\end{equation}
Stability of a stationary point $\bPhi_{\ast}$ is hence determined
by the Jacobian of the vector field $\bG(\by^{(0)}_{\ast})$,
which depends on the averaged terms $\mathcal{T}_{i}$
and the bending potential $\Vbend^{(2)}$.

In this section we concentrate on
\begin{equation}
  \Vbend^{(2)}(\by)\equiv 0
\end{equation}
to clarify the dynamically induced stability by the averaged terms $\mathcal{T}_{i}$.
According to Eq.~\eqref{eq:Gi},
$E^{(2)}$ is an overall factor of the function $G_{i}$
when the bending potential is absent,
and stability does not depend on $E^{(2)}$.
We come back to the chain-like system
\begin{equation}
  \mK_{l} = {\rm diag}(k_{1},\cdots,k_{N-1}) \in {\rm Diag}(N-1).
\end{equation}
Further, we restrict ourselves to the uniform setting
\begin{equation}
  \label{eq:uniform-setting}
  m_{i} = m,~ k_{j}=k, ~ l_{j,\ast}=l_{\ast}
  \quad
  (1\leq i\leq N; 1\leq j\leq N-1)
\end{equation}
and to the one-dimensional conformations introduced in Sec.~\ref{sec:1Dconf}.

\subsection{One-dimensional conformations}
\label{sec:1Dconf}

We introduce the one-dimensional conformations
whose set is denoted by
\begin{equation}
  \mathcal{C}^{1} = \left\{ (\phi_{1},\cdots,\phi_{N-2})
    ~|~ \phi_{i}\in \{0,\pi\} ~(i=1,\cdots,N-2) \right\}.
\end{equation}
A conformation in $\mathcal{C}^{1}$ is stationary
as proven in Appendix \ref{sec:App-Theory-1Dconf}.
Appearance of the bending potential forbids $\phi_{i}=\pi$ in general
to avoid collision between beads, but we accept $\phi_{i}=\pi$ in this section
to discuss the simplest case.
Later we will perform numerical simulations under appearance of a bending potential
which forbids $\phi_{i}=\pi$.

\begin{figure}[h]
  \centering
  \includegraphics[width=8cm]{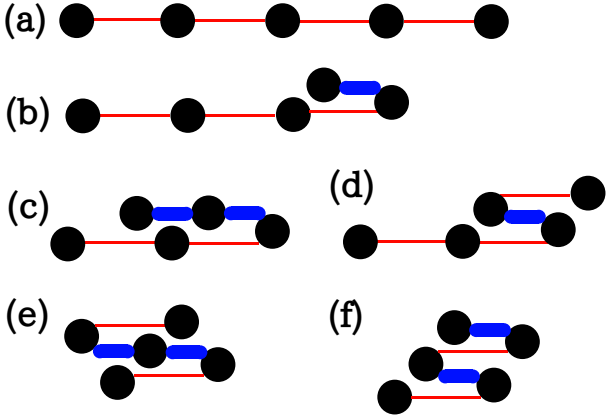}
  \caption{Illustration of the one-dimensional conformations for $N=5$.
    The stabilizing mode is characterized by the two types of bonds:
    A red thin (a blue thick) bond represents that
    the spring is longer (shorter) than the natural length.
    Conformation symbols are
    (a) $\bc=(1,1,1)$, (b) $\bc=(1,1,-1)$,
    (c) $\bc=(1,-1,1)$, (d) $\bc=(1,-1,-1)$,
    (e) $\bc=(-1,1,-1)$, and (f) $\bc=(-1,-1,-1)$.
  }
  \label{fig:StableFormsN5}
\end{figure}

A conformation in $\mathcal{C}^{1}$ is symbolized
by a sequence of $1$ and $-1$:
$1$ represents the straight joint ($\phi=0$),
and $-1$ the fully bent joint ($\phi=\pi$).
The conformation symbol is denoted by $\bc=(c_{1},\cdots,c_{N-2})$.
We identify two symmetric conformations like $(1,-1,-1)$ and $(-1,-1,1)$,
because each of which is mapped to the other
by changing the starting end of the chain.
All the possible one-dimensional conformations for $N=5$
are illustrated in Fig.~\ref{fig:StableFormsN5}
with their conformation symbols.

\subsection{Stability}
Let the point $\bPhi_{\ast}$ be stationary.
Stability of this stationary point is obtained from the eigenvalues
of the Jacobian matrix for Eq.~\eqref{eq:EOM-in-normalform},
\begin{equation}
  \mJ(\bPhi_{\ast}) =
  \begin{pmatrix}
    \mO & \mE \\
    -D \bG(\by_{\ast}^{(0)}) & \mO \\
  \end{pmatrix},
\end{equation}
where
\begin{equation}
  D\bG =
  \begin{pmatrix}
    \dfracp{G_{1}}{\phi_{1}} & \cdots & \dfracp{G_{1}}{\phi_{N-1}} \\
    \vdots & \ddots & \vdots \\
    \dfracp{G_{N-1}}{\phi_{1}} & \cdots & \dfracp{G_{N-1}}{\phi_{N-1}} \\
  \end{pmatrix}
\end{equation}
is the Jacobian matrix of the vector field $\bG(\by)$.
We do not take the derivatives with respect to the variables $\bl$,
because Eq.~\eqref{eq:EOM-in-normalform} includes
only the constant lengths $\bl_{\ast}$.

\begin{figure}[h]
  \centering
  \includegraphics[width=7cm]{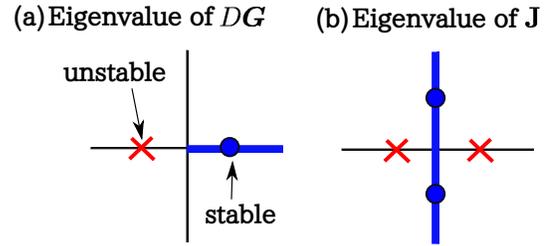}
  \caption{Eigenvalues of $D\bG$ and $\mJ$ on the complex plane.
    The blue circle in the panel (a) produces the two blue circles in the panel (b),
    and the same for the red crosses. 
    A conformation is stable if and only if all the eigenvalues
    are on the blue thick line.}
  \label{fig:EigenvalueDGJ}
\end{figure}

Let us assume that the matrix $D\bG(\by_{\ast}^{(0)})$ is diagonalizable
and has the eigenvalues $g_{1},\cdots,g_{N-1}$.
One eigenvalue, denoted by $g_{N-1}$,
should be zero from the rotational symmetry,
and we remove it from the stability criterion.
The nontrivial eigenvalues of $\mJ(\bPhi_{\ast})$ are
$\pm\sqrt{-g_{1}}, \cdots, \pm\sqrt{-g_{N-2}}$.
See Appendix \ref{sec:extended-eigenvalues} for a proof
and Fig.~\ref{fig:EigenvalueDGJ} for a schematic explanation
of the relation between the eigenvalues of $D\bG$ and $\mJ$.

An eigenvalue $g_{i}$ is called a stable eigenvalue if $g_{i}\in (0,\infty)$,
and a zeroeigenvalue if $g_{i}=0$,
and an unstable eigenvalue if $g_{i}\not\in [0,\infty)$.
The conformation represented by $\bphi^{(0)}_{\ast}$ is (neutrally) stable
\cite{holm-etal-85} if and only if there is no unstable eigenvalue.

\subsection{Mode selectivity of dynamical stabilization}
\label{sec:Theory-pure-uni}

We first excite only one normal mode of the springs:
All the diagonal elements of $\mN$ are zero except for one element.
Stability of the one-dimensional conformations is summarized in Table \ref{tab:Stability}
with dependency on the excited normal mode,
where the normal modes are numbered in the ascending order of the eigenfrequencies
around the considering conformation,
as mentioned after Eq.~\eqref{eq:Hypothesis}.
Stability is symbolized by S, Z, and U,
and the number after S (Z, U) represents the number of the stable
(zero, unstable) eigenvalues of $D\bG$, whose sum is $N-2$.
The symbol is omitted if the number of corresponding eigenvalues is zero.
For instance, the symbol S2U1 represents that the conformation
has $2$ stable eigenvalues and $1$ unstable eigenvalue,
and the conformation is unstable.

\begin{widetext}

\begin{table}
  \centering
  \begin{tabular}{l|l|lllll}
    \hline
    $N$ & Conformation & \multicolumn{5}{l}{Stability / Square of eigenfrequency} \\
    \cline{3-7} 
    & $\bc$ & Mode-$1$ & Mode-$2$ & Mode-$3$ & Mode-$4$ & Mode-$5$ \\
    \hline
    $3$ &  & S1 / 1 & U1 / 3 \\
    \cline{3-7} 
    & $(1)$ & $(+,+)$ & $(+,-)$ \\
    & $(-1)$ & $(+,-)$ & $(+,+)$ \\
    \hline
    $4$ & & S2 / 0.585786 & Z2 / 2 & U2 / 3.41421 \\
    \cline{3-7} 
        & $(1,1)$ & $(+,+,+)$ & $(+,0,-)$ & $(+,-,+)$ \\
        & $(1,-1)$ & $(+,+,-)$ & $(+,0,+)$ & $(+,-,-)$ \\
        & $(-1,-1)$ & $(+,-,+)$ & $(+,0,-)$ & $(+,+,+)$ \\
    \hline
    $5$ & & S3 / 0.381966 & S2U1 / 1.38197 & S1U2 / 2.61803 & U3 / 3.61803 \\
    \cline{3-7} 
        & $(1,1,1)$ & $(+,+,+,+)$ & $(+,+,-,-)$ & $(+,-,-,+)$ & $(+,-,+,-)$ \\
        & $(1,1,-1)$ & $(+,+,+,-)$ & $(+,+,-,+)$ & $(+,-,-,-)$ & $(+,-,+,+)$ \\
        & $(1,-1,1)$ & $(+,+,-,-)$ & $(+,+,+,+)$ & $(+,-,+,-)$ & $(+,-,-,+)$ \\
        & $(1,-1,-1)$ & $(+,+,-,+)$ & $(+,+,+,-)$ & $(+,-,+,+)$ & $(+,-,-,-)$ \\
        & $(-1,1,-1)$ & $(+,-,-,+)$ & $(+,-,+,-)$ & $(+,+,+,+)$ & $(+,+,-,-)$ \\
        & $(-1,-1,-1)$ & $(+,-,+,-)$ & $(+,-,-,+)$ & $(+,+,-,-)$ & $(+,+,+,+)$ \\
    \hline
    $6$ & & S4 / 0.267949 & S2Z2 / 1 & Z4 / 2 & Z2U2 / 3 & U4 / 3.73205 \\
    \cline{3-7} 
        & $(1,1,1,1)$ & $(+,+,+,+,+)$ ~ & $(+,+,0,-,-)$ ~ & $(+,0,-,0,+)$ ~ & $(+,-,0,+,-)$ ~ & $(+,-,+,-,+)$ \\
        & $(1,1,1,-1)$ & $(+,+,+,+,-)$ & $(+,+,0,-,+)$ & $(+,0,-,0,-)$ & $(+,-,0,+,+)$ & $(+,-,+,-,-)$ \\
        & $(1,1,-1,1)$ & $(+,+,+,-,-)$ & $(+,+,0,+,+)$ & $(+,0,-,0,-)$ & $(+,-,0,-,+)$ & $(+,-,+,+,-)$ \\
        & $(1,1,-1,-1)$ & $(+,+,+,-,+)$ & $(+,+,0,+,-)$ & $(+,0,-,0,+)$ & $(+,-,0,-,-)$ & $(+,-,+,+,+)$ \\
        & $(1,-1,1,1)$ & $(+,+,-,-,-)$ & $(+,+,0,+,+)$ & $(+,0,+,0,-)$ & $(+,-,0,-,+)$ & $(+,-,-,+,-)$ \\
        & $(1,-1,1,-1)$ & $(+,+,-,-,+)$ & $(+,+,0,+,-)$ & $(+,0,+,0,+)$ & $(+,-,0,-,-)$ & $(+,-,-,+,+)$ \\
        & $(1,-1,-1,1)$ & $(+,+,-,+,+)$ & $(+,+,0,-,-)$ & $(+,0,+,0,+)$ & $(+,-,0,+,-)$ & $(+,-,-,-,+)$ \\
        & $(1,-1,-1,-1)$ & $(+,+,-,+,-)$ & $(+,+,0,-,+)$ & $(+,0,+,0,-)$ & $(+,-,0,+,+)$ & $(+,-,-,-,-)$ \\
        & $(-1,1,1,-1)$ & $(+,-,-,-,+)$ & $(+,-,0,+,-)$ & $(+,0,+,0,+)$ & $(+,+,0,-,-)$ & $(+,+,-,+,+)$ \\
        & $(-1,1,-1,-1)$ & $(+,-,-,+,-)$ & $(+,-,0,-,+)$ & $(+,0,+,0,-)$ & $(+,+,0,+,+)$ & $(+,+,-,-,-)$ \\
        & $(-1,-1,-1,-1)$ & $(+,-,+,-,+)$ & $(+,-,0,+,-)$ & $(+,0,-,0,+)$ & $(+,+,0,-,-)$ & $(+,+,+,+,+)$ \\
    \hline
  \end{tabular}
  \caption{Stability of the one-dimensional conformations
    in the $N$-body chain-like bead-spring model
    under the equal masses and the identical springs.
    The second column from the left represents the conformation symbol $\bc$.
    The mode number is defined in the ascending order of the eigenfrequencies.
    Stability of a conformation with a given mode
    is indicated by S (stable), Z(zero), and U (unstable),
    and the number after S (Z, U) represents the number of
    stable (zero, unstable) eigenvalues of $D\bG$.
    After the slash, the value of $(m/k)\omega_{j}^{2}$ is shown,
    where $\omega_{j}$ is the eigenfrequency of the mode.
    Sequences of $+,0$, and $-$ represent the eigenmode symbol $\bs$.
    See the text for the definitions of $\bc$ and $\bs$.
  }
  \label{tab:Stability}
\end{table}

\end{widetext}

The $(N-1)$-dimensional eigenvector of a normal mode is characterized
by a sequence of $+, 0$, and $-$.
The symbol $+~(0, -)$ represents that the corresponding spring
is longer than (equal to, shorter than) the natural length.
That is, for $N=3$, the eigenmode $(+,+)$ implies that the two springs
are initially longer than the natural length and the mode is the in-phase mode.
Similarly, the eigenmode $(+,-)$ represents the antiphase mode.
We denote the eigenmode symbol by $\bs=(s_{1},\cdots,s_{N-1})$.

We have two observations in Table \ref{tab:Stability}.
First, each conformation is stabilized by the lowest eigenfrequency mode of the springs.
The number of unstable directions increases as the eigenfrequency gets larger.
Second, the stabilizing eigenmode is obtained by pulling the left (right) end
of the chain to the left (right) as illustrated in Fig.~\ref{fig:StableFormsN5}.

Stability analysis can be extended to mixed modes.
Analyses for $N=3,4$ and $5$ suggest that the dynamical stabilization
is ubiquitous in a larger system having multimode excitation.
Indeed, the dynamical stabilization is realized with an approximate probability
of $0.8$ up to $N=5$, whereas higher eigenfrequency modes contribute
to destabilization. See Appendix \ref{sec:Theory-mixed-uni}.

\section{Numerical tests}
\label{sec:Numerics}

We demonstrate dynamical stabilization and destabilization
through numerical simulations of the system
under the uniform setting \eqref{eq:uniform-setting} with
\begin{equation}
  \label{eq:uniform-setting-values}
  m=1, \quad k=10, \quad l_{\ast}=1. 
\end{equation}
We use the Hamiltonian written in the Cartesian coordinate
to use an explicit fourth-order symplectic integrator \cite{yoshida-90}
with the time step $\Delta t=10^{-3}$.
The Hamiltonian associated with the Lagrangian \eqref{eq:L-in-r} is
\begin{equation}
  H(\br, \bp) = \dfrac{1}{2m} \sum_{i=1}^{N} \norm{\bp_{i}}^{2}
  + \Vspring(\br) + \Vbend(\br),
\end{equation}
and the canonical equations of motion are
\begin{equation}
  \dfracd{\br_{i}}{t} = \dfracp{H}{\bp_{i}},
  \quad
  \dfracd{\bp_{i}}{t} = - \dfracp{H}{\br_{i}},
  \quad
  (i=1,\cdots,N).
\end{equation}

\subsection{System setting}

The theory includes the spring potential $\Vspring$ up to the quadratic order,
and we use the linear springs.
The spring potential $\Vspring$ is defined in Eq.\eqref{eq:Vspring},
and each spring $V_{i}$ is
\begin{equation}
  V_{i}(\br) = \dfrac{k}{2} ( \norm{\br_{i+1}-\br_{i}} - l_{\ast})^{2},
  \quad
  (i=1,\cdots,N-1).
\end{equation}

The bending potential $\Vbend$ is
\begin{equation}
  \label{eq:Vbend}
  \Vbend(\br) = \sum_{i<j} U_{\rm LJ}(\norm{\br_{j}-\br_{i}}),
\end{equation}
and $U_{\rm LJ}$ is the Lennard-Jones potential
\begin{equation}
  U_{\rm LJ}(r) = 4 \epsilon_{\rm LJ} \left[
    \left( \dfrac{\sigma}{r} \right)^{12} - \left( \dfrac{\sigma}{r} \right)^{6}
  \right].
\end{equation}
The parameter $\epsilon_{\rm LJ}$ is of $O(\epsilon^{2})$, namely
\begin{equation}
  \epsilon_{\rm LJ} = \epsilon_{0} \epsilon^{2},
  \quad
  \epsilon_{0} = O(\epsilon^{0}),
\end{equation}
to satisfy $\Vbend=O(\epsilon^{2})$.
We fix $\epsilon_{\rm LJ}$ and $\sigma$ as
\begin{equation}
  \label{eq:epsilonLJ-sigma}
  \epsilon_{\rm LJ}= 10^{-4}, \quad \sigma=1.
\end{equation}

\begin{figure}
  \centering
  \includegraphics[width=8cm]{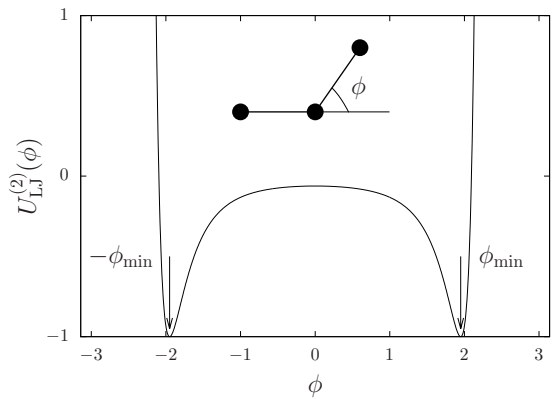}
  \caption{The second-order Lennard-Jones potential \eqref{eq:ULJ2}.
    The two arrows mark the minimum points $\pm\phi_{\rm min}$ \eqref{eq:phimin}.
    $\epsilon_{0}=1, \sigma=1$, and $\l_{\ast}=1$.
    The definition of the bending angle $\phi$ is schematically represented
    at the top center. }
  \label{fig:ULJ2}
\end{figure}

We may expect that the main contribution to the bending energy
comes from pairs of second nearest beads,
since the Lennard-Jones potential does not depend on the bending angles
for a pair of nearest beads.
For a second nearest pair with the bending angle $\phi$,
the second-order Lennard-Jones potential is
\begin{equation}
  \label{eq:ULJ2}
  U_{\rm LJ}^{(2)}(\phi) = 4\epsilon_{0} \left\{
    \left[ \dfrac{\sigma^{2}}{2l_{\ast}^{2}(1+\cos\phi)} \right]^{6}
    - \left[ \dfrac{\sigma^{2}}{2l_{\ast}^{2}(1+\cos\phi)} \right]^{3}
  \right\}.
\end{equation}
It takes the minimum value $-\epsilon_{0}$ at $\pm\phi_{\rm min}$, where
\begin{equation}
  \label{eq:phimin}
  \phi_{\rm min}
  = \left| \cos^{-1} \left[ \left( \dfrac{\sigma}{2^{1/3}l_{\ast}} \right)^{2} - 1 \right] \right|
  \simeq 1.94985.
\end{equation}
See Fig.~\ref{fig:ULJ2}.

It is worth noting that for $N=3$ we can construct the effective potential
\cite{yamaguchi-etal-22},
which is useful to understand stability of a conformation graphically.
Examples are exhibited in Appendix \ref{sec:effective-potential}.

\subsection{Initial conditions}
\label{sec:Numerics-ini}

\begin{figure}[b]
  \centering
  \includegraphics[width=8cm]{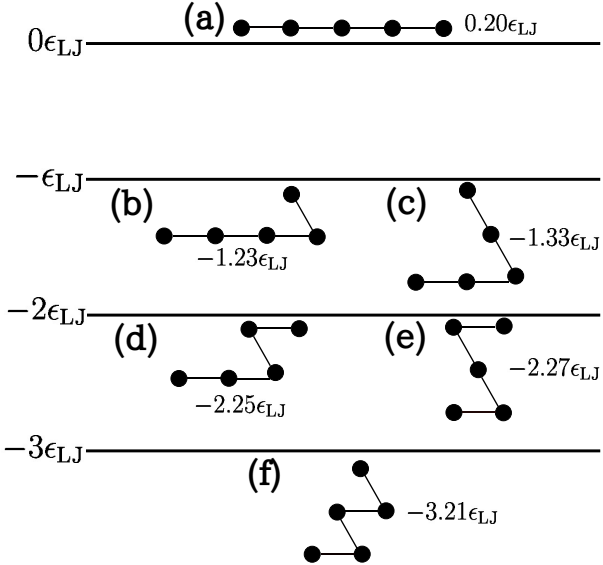}
  \caption{Illustration of the possible initial conformations
    and their approximated bending potential energy $\Vbend$ for $N=5$.
    Conformation symbols are
    (a) $\bc=(1,1,1)$, (b) $\bc=(1,1,-1)$,
    (c) $\bc=(1,-1,1)$, (d) $\bc=(1,-1,-1)$,
    (e) $\bc=(-1,1,-1)$, and (f) $\bc=(-1,-1,-1)$.
    The bending angle $\pi$ has been replaced with $\pm\phi_{\rm min}$
    from Fig.~\ref{fig:StableFormsN5}.
  }
  \label{fig:InitialPosition}
\end{figure}

The initial positions are given in the following three steps.
First, we select a reference conformation from $\mathcal{C}^{1}$
and put all the beads on the $x$-axis.
Second, we replace the bending angle $\phi=\pi$ with $\phi=\pm\phi_{\rm min}$
to avoid collision of beads.
The possible initial conformations for $N=5$
are illustrated in Fig.~\ref{fig:InitialPosition}.
Third,  we modify the lengths of the springs from the natural length to
$\bl=\bl_{\ast}+\epsilon \bl^{(1)}$,
where $\bl^{(1)}$ is determined so as to excite normal modes
in a desired manner approximately.
Precise settings of the initial positions are described
in Appendix \ref{sec:initial-position}.

The initial values of the momenta are set as follows.
For the $x$-direction, the initial values of the momenta are zero.
For the $y$-direction, they are randomly drawn
from the uniform distribution on the interval $[-10^{-3}, 10^{-3}]$
as perturbation to observe stability of the given conformation.

\subsection{Dynamical stabilization and destabilization}
\label{sec:DIC}

\begin{figure}[tb]
  \centering
  \includegraphics[width=8.5cm]{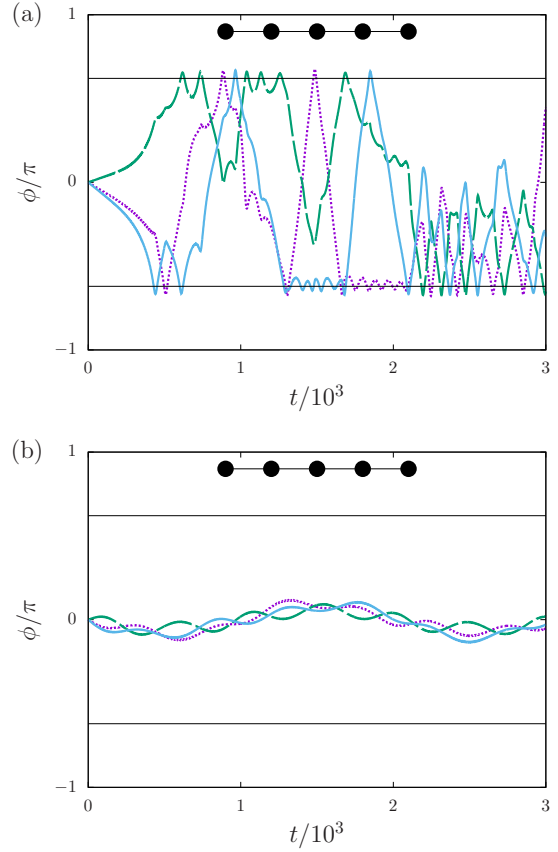}
  \caption{Temporal evolution of the bending angles.
    $N=5$. Conformation is $\bc=(1,1,1)$, schematically represented
    at the top of each panel, and the mode-$1$ is excited.
    (a) $\epsilon=0.03$. $E\simeq -0.149\epsilon_{\rm LJ}$
    (b) $\epsilon=0.04$. $E\simeq 0.070\epsilon_{\rm LJ}$.
    $\epsilon_{\rm LJ}=10^{-4}$.
    $\phi_{1}(t)$ (purple dotted line),  $\phi_{2}(t)$ (green broken line),
    and $\phi_{3}(t)$ (blue solid line).
    Two black horizontal lines are shown at $\pm\phi_{\min}$.
    }
  \label{fig:StraightConfStability}
\end{figure}

We examine dynamical stabilization of high energy conformations.
The first example is the straight conformation,
which is illustrated in Fig.~\ref{fig:InitialPosition}(a).
Temporal evolution of $\phi_{i}(t)$ is exhibited in Fig.~\ref{fig:StraightConfStability}
by exciting the mode-$1$ (the stabilization mode)
and varying the amplitude $\epsilon$ of the mode.
For $\epsilon=0.03$ the bending angles oscillate between
the two points $\pm\phi_{\rm min}$.
However, for $\epsilon=0.04$ which provides larger $E^{(2)}$,
the straight conformation is stabilized and the bending angles stay around $0$.
This observation is consistent with the expression of $G_{i}$ \eqref{eq:Gi},
because larger $E^{(2)}$ enhances contribution from the averaged term $\mathcal{T}_{s}$.

\begin{figure}[tb]
  \centering
  \includegraphics[width=8.5cm]{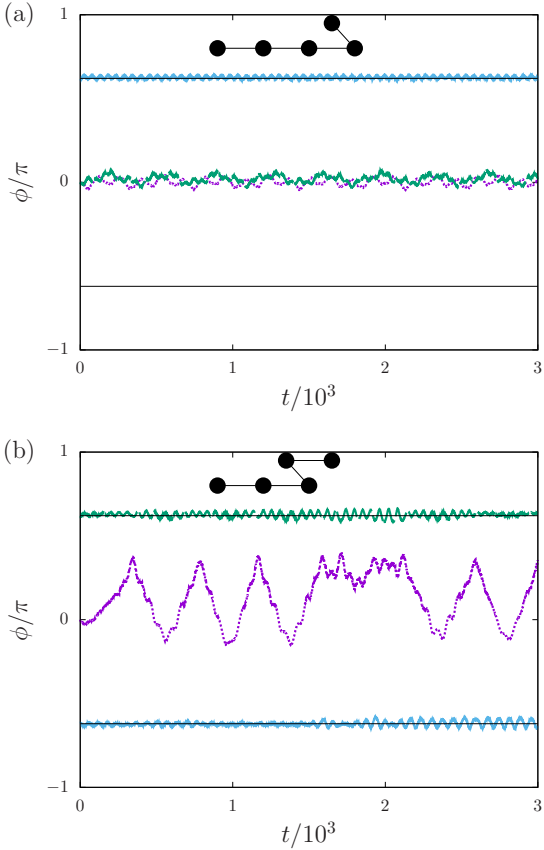}
  \caption{Temporal evolution of the bending angles.
    $N=5$. The mode-$1$ is excited.
    (a) Conformation $\bc=(1,1,-1)$, $\epsilon=0.12$, and $E\simeq 5.22\epsilon_{\rm LJ}$.
    (b) Conformation $\bc=(1,-1,-1)$, $\epsilon=0.10$, and $E\simeq 2.45\epsilon_{\rm LJ}$.
    $\epsilon_{\rm LJ}=10^{-4}$.
    The conformation is schematically represented at the top of each panel.
    $\phi_{1}(t)$ (purple dotted line),  $\phi_{2}(t)$ (green broken line),
    and $\phi_{3}(t)$ (blue solid line).
    Two black horizontal lines are shown at $\pm\phi_{\min}$.
    }
  \label{fig:PartiallyStraightConfStability}
\end{figure}

The stabilization by the mode-$1$ is also realized for partially straight conformations.
For $N=5,$ temporal evolution of $\phi_{i}~(i=1,2,3)$ are reported in
Fig.~\ref{fig:PartiallyStraightConfStability}
for conformations symbolized by $\bc=(1,1,-1)$
[Fig.~\ref{fig:InitialPosition}(b)]
and $\bc=(1,-1,-1)$ [Fig.~\ref{fig:InitialPosition}(d)].
The bending angles stay around the initial values.

\begin{figure}[tb]
  \centering
  \includegraphics[width=8.5cm]{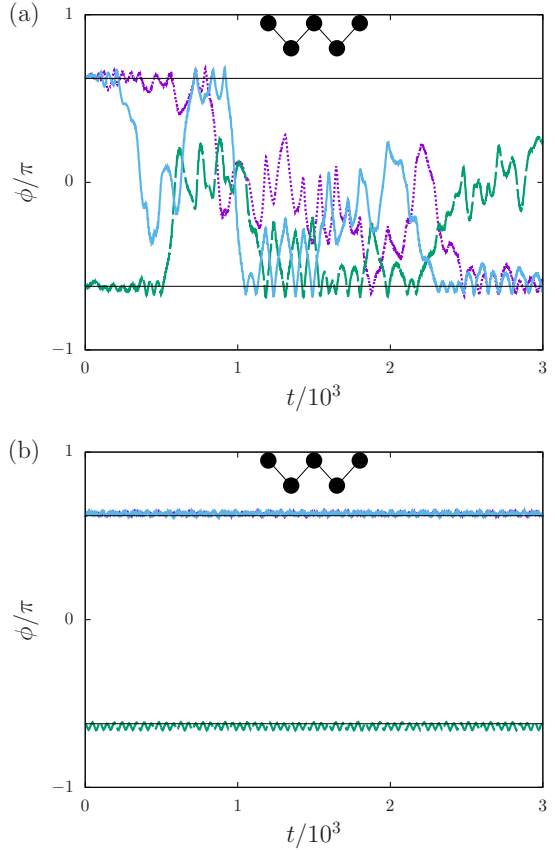}
  \caption{Excited mode dependence of the bent conformation $\bc=(-1,-1,-1)$,
    schematically represented at the top of each panel.
    $N=5$. $\epsilon=0.16$.
    (a) The mode-$2$ is excited and $E\simeq 9.6 \epsilon_{\rm LJ}$.
    (b) The mode-$1$ is excited and $E\simeq 10.2\epsilon_{\rm LJ}$.
    $\epsilon_{\rm LJ}=10^{-4}$.
    $\phi_{1}(t)$ (purple dotted line),  $\phi_{2}(t)$ (green broken line),
    and $\phi_{3}(t)$ (blue solid line).
    Two lines are almost collapsed around $\phi_{\rm min}$ in the panel (b).
    Two black horizontal lines are shown at $\pm\phi_{\min}$.
    }
  \label{fig:BendConfDestability}
\end{figure}

Finally, we demonstrate destabilization of the bent conformation
$\bc=(-1,-1,-1)$ [Fig.~\ref{fig:InitialPosition}(f)],
which is stable if dynamical instability does not kick in.
The destabilization is realized by the mode-$2$ for instance as shown
in Fig.~\ref{fig:BendConfDestability}(a),
which is consistent with Table~\ref{tab:Stability},
although larger spring energy is necessary to destabilize the bent conformation.
We stress that the destabilization is not induced only by largeness of energy,
because the bent conformation is not destabilized by the mode-$1$
as shown in Fig.~\ref{fig:BendConfDestability}(b),
while the values of energy are almost equal between the two cases.

\section{Summary}
\label{sec:Summary}

We have studied the dynamically induced conformation (DIC)
in $N$-body chain-like bead-spring models.
We have extended a theory, which is developed for $N=3$ in a previous work
\cite{yamaguchi-etal-22}, to a general $N$.
The theory predicts that the dynamical stability depends on the excited
normal modes of the springs and on the value of energy.

As the simplest case we have studied a system without the bending potential
to clearly exhibit dynamical effects.
Concentrating on the so-called one-dimensional conformations, which are stationary,
We have investigated the mode dependent stability up to $N=6$
under the condition of the equal masses and the identical springs.
A simple rule of the mode dependency has been discovered:
A conformation is stabilized by exciting the lowest eigenfrequency mode,
and destabilization emerges as the eigenfrequency of the exited pure normal mode
becomes higher.

We stress that DIC is ubiquitous.
The theory is also applicable for mixed modes,
and the stabilization of a conformation is realized
with an approximate probability of $0.8$ up to $N=5$,
when we randomly choose a mixed mode.
The probability $0.8$ is notable, because, among four normal modes in $N=5$,
only one mode contributes to the stabilization
and the other three modes contribute to the destabilization.
Moreover, the uniform setting of the equal masses and the identical springs
is not essential for DIC \cite{yamaguchi-etal-22}.

The stabilization and destabilization of conformations
have been demonstrated numerically in a system having the bending potential
consisting of the Lennard-Jones potentials for each pair of beads.
As the theory predicts, any conformation can be stabilized by exciting the lowest
eigenfrequency mode which depends on the conformation,
whereas a straight joint corresponds to the local maximum of a Lennard-Jones potential.
Destabilization of the fully bend conformation,
which corresponds to a local minimum point of the bending potential,
has been also demonstrated by exciting a higher eigenfrequency mode.

We note that excitation of a normal mode is a nonequilibrium phenomenon,
because the law of equipartition of energy holds among the normal modes
in thermal equilibrium.
Nevertheless, separation of the two timescales suggests
that importance of DIC survives in a long time by the Boltzmann-Jeans conjecture
\cite{boltzmann-95,jeans-03,jeans-05,landau-teller-36,benettin-galgani-giorgilli-89,baldin-benettin-91}.
An important message of DIC is that the conformation is not determined
by the bending potential only,
and we have to input the dynamical (de)stabilization.
This message sheds light on a new aspect of conformation and conformation change.

\acknowledgements
The author thanks T. Yanagita, T. Konishi, and M. Toda for valuable discussions.
The author acknowledges the support of JSPS KAKENHI Grant Numbers 16K05472 and 21K03402.

\appendix

\section{Lagrangian in the internal coordinates}
\label{sec:L-in-y}

We rewrite the Lagrangian Eq.~\eqref{eq:L-in-r} into the internal coordinates
through three changes of variables.
We mainly consider modifications of the kinetic energy
\begin{equation}
  K(\dot{\br}) = \dfrac{1}{2} \dot{\br}^{\rm T} \mM \dot{\br},
\end{equation}
where
\begin{equation}
  \mM = {\rm diag}(m_{1},\cdots,m_{N}) \in {\rm Diag}(N)
\end{equation}
and the symbol ${\rm Diag}(n)$ represents the set of the real diagonal matrices
of size $n$.

The first change of variables is
\begin{equation}
    \begin{pmatrix}
      \bq_{1} \\ \vdots \\ \bq_{N}
    \end{pmatrix}
    = \mS_{M}
    \begin{pmatrix}
      \br_{1} \\ \vdots \\ \br_{N}
    \end{pmatrix}
\end{equation}
where the matrix $\mS_{M}\in{\rm Mat}(N)$ is
\begin{equation}
  \mS_{M} =
  \begin{pmatrix}
    -1 & 1 & 0 & \cdots & 0 & 0 \\
    0 & -1 & 1 & \ddots & 0 & 0 \\
    0 & 0 & -1 & \ddots & 0 & 0 \\
    \vdots & \ddots & \ddots & \ddots & \ddots & \vdots \\
    0 & 0 & 0 & \cdots & -1 & 1 \\
    m_{1}/M & m_{2}/M & m_{3}/M & \cdots & m_{N-1}/M & m_{N}/M \\
  \end{pmatrix}
\end{equation}
and 
\begin{equation}
  M = \Tr \mM = \sum_{i=1}^{N} m_{i}.
\end{equation}
In the new variables $\bq_{i}$,
the kinetic energy $K$ is
\begin{equation}
  \label{eq:K-in-q}
  K(\dot{\bq}) = \dfrac{1}{2} \sum_{i,j=1}^{N-1} A^{ij} \dot{\bq}_{i} \cdot \dot{\bq}_{j}
  + \dfrac{\mu}{2} \norm{\dot{\bq}_{N}}^{2}.
\end{equation}
The symmetric constant matrix $\mA\in{\rm Mat}(N-1)$ is defined by
\begin{equation}
  \mS_{M}^{-{\rm T}}\mM \mS_{M}^{-1}
  =
  \begin{pmatrix}
    \mA & \bzero \\
    \bzero^{\rm T} & n^{2} \\
  \end{pmatrix},
\end{equation}
where the superscript $-$T represents transposition of the inverse matrix
and $\bzero$ is the zero column vector.
The variable $\bq_{N}$ is a cyclic coordinate
corresponding to the total momentum conservation
due to the translational symmetry.
We set the total momentum as zero,
and we drop the last term of the right-hand side of Eq.~\eqref{eq:K-in-q}.

The second change of variables introduce the polar coordinates.
Denoting $\bq_{i}=(q_{xi},q_{yi})\in\R^{2}$, we introduce $l_{i}$ and $\theta_{i}$ by
\begin{equation}
  q_{xi} = l_{i} \cos\theta_{i},
  \quad
  q_{yi} = l_{i} \sin\theta_{i}.
\end{equation}
In the vector form, the polar coordinates are expressed by
\begin{equation}
  \bq_{i} = l_{i} \be_{li},
  \quad
  \dot{\bq}_{i} = \dot{l}_{i} \be_{li} + l_{i} \dot{\theta}_{i} \be_{\theta i},
\end{equation}
where $\be_{li}$ and $\be_{\theta i}$ are the unit vectors
of the radial and the angle directions, respectively.
The polar coordinates rewrite the spring potential as $\Vspring(\bl)$,
and the kinetic energy $K$ as
\begin{equation}
  K
  = \dfrac{1}{2}
  \begin{pmatrix}
    \dot{\bl}^{\rm T} & \dot{\btheta}^{\rm T}
  \end{pmatrix}
  \begin{pmatrix}
    \mA_{C}(\btheta) & \mA_{S}(\btheta) \mL(\bl) \\
    - \mL(\bl) \mA_{S}(\btheta) & \mL(\bl) \mA_{C}(\btheta) \mL(\bl) \\
  \end{pmatrix}
  \begin{pmatrix}
    \dot{\bl} \\ \dot{\btheta} \\
  \end{pmatrix}.
\end{equation}
The diagonal matrix $\mL$ is defined by
\begin{equation}
  \mL(\bl) = {\rm diag}(l_{1},\cdots,l_{N-1}) \in {\rm Diag}(N-1).
\end{equation}
The symmetric matrix $\mA_{C}$ and the antisymmetric matrix $\mA_{S}$
are defined by
\begin{equation}
  A_{C}^{ij}(\btheta) = A^{ij} \cos(\theta_{i}-\theta_{j}),
  \quad
  A_{S}^{ij}(\btheta) = A^{ij} \sin(\theta_{i}-\theta_{j}).
\end{equation}

The third change of variables introduces the bending angles $\bphi$ as
\begin{equation}
  \label{eq:theta2phi}
  \begin{pmatrix}
    \phi_{1} \\ \vdots \\ \phi_{N-1} \\
  \end{pmatrix}
  = \mS
  \begin{pmatrix}
    \theta_{1} \\ \vdots \\ \theta_{N-1} \\
  \end{pmatrix},
\end{equation}
where the constant matrix $\mS$ is
\begin{equation}
  \mS =
  \begin{pmatrix}
    -1 & 1 & 0 & \cdots & 0 & 0 \\
    0 & -1 & 1 & \ddots & 0 & 0 \\
    0 & 0 & -1 & \ddots & 0 & 0 \\
    \vdots & \ddots & \ddots & \ddots & \ddots & \vdots \\
    0 & 0 & 0 & \cdots & -1 & 1 \\
    \frac{1}{N-1} & \frac{1}{N-1} & \frac{1}{N-1} & \cdots & \frac{1}{N-1} & \frac{1}{N-1} \\
  \end{pmatrix}
  \in {\rm Mat}(N-1).
\end{equation}
The bending angles are from $\phi_{1}$ to $\phi_{N-2}$,
and we added an additional angle $\phi_{N-1}$ for later convenience.
Performing the change of variables of Eq.~\eqref{eq:theta2phi},
the kinetic energy is modified to 
\begin{equation}
  \label{eq:K-in-y}
  K(\by,\dot{\by})
  = \dfrac{1}{2} \dot{\by}^{\rm T} \mB(\by) \dot{\by},
\end{equation}
where the matrix $\mB$ is
\begin{equation}
  \mB(\by) =
  \begin{pmatrix}
    \mB_{ll}(\bphi) & \mB_{l\phi}(\by) \\
    \mB_{\phi l}(\by) & \mB_{\phi\phi}(\by) \\
  \end{pmatrix}
  \in \MatN
\end{equation}
and the size-$(N-1)$  submatrices are defined by
\begin{equation}
  \label{eq:B-submatrices}
  \begin{pmatrix}
    \mB_{ll} & \mB_{l\phi} \\
    \mB_{\phi l} & \mB_{\phi\phi}
  \end{pmatrix}
  =
  \begin{pmatrix}
    \mA_{C}(\bphi) & \mA_{S}(\bphi) \mL \mS^{-1} \\
    - \mS^{-{\rm T}} \mL \mA_{S}(\bphi) & \mS^{-{\rm T}} \mL \mA_{C}(\bphi) \mL \mS^{-1} \\
  \end{pmatrix}.
\end{equation}
The $(i,j)$ elements of the matrices
$\mA_{C}(\bphi),\mA_{S}(\bphi)\in{\rm Mat}(N-1)$ are written respectively as
\begin{equation}
  \label{eq:ACAS-in-phi}
  A_{C}^{ij}(\bphi) = A^{ij} \cos\phi_{i,j},
  \quad
  A_{S}^{ij}(\bphi) = A^{ij} \sin\phi_{i,j},
\end{equation}
where
\begin{equation}
  \phi_{i,j} = \theta_{i} - \theta_{j}
  = \left\{
    \begin{array}{ll}
      \phi_{i-1} + \cdots + \phi_{j} & (i>j), \\
      0 & (i=j), \\
      -(\phi_{j-1}+\cdots+\phi_{i}) & (i<j). \\
    \end{array}
  \right.
\end{equation}
The variable $\phi_{N-1}$ does not appear in $\mB(\by)$.

After the three changes of variables,
we obtain the Lagrangian in the internal coordinates as Eq.~\eqref{eq:L-in-y}.
The variable $\phi_{N-1}$ is a cyclic coordinate
corresponding to the rotational symmetry.
We keep it for later convenience of computations.

\section{Ordering of the bending potential}
\label{sec:ordering-bending-potential}

Since $\dot{\by},\ddot{\by}=O(\epsilon)$,
the zeroth order equations of motion are
\begin{equation}
  \left( \dfracp{V}{y_{\alpha}}(\by) \right)^{(0)} = 0,
  \quad
  (\alpha=1,\cdots,2N-2)
\end{equation}
which implies
\begin{equation}
  \left( \dfracp{V}{y_{\alpha}}(\by) \right)^{(0)}
  = \dfracp{\Vbend^{(0)}}{y_{\alpha}}(\by^{(0)}) \equiv 0,
  \quad
  (\alpha=1,\cdots,2N-2).
\end{equation}
Remembering $\by^{(0)}=(\bl_{\ast},\bphi^{(0)}(t_{1}))^{\rm T}$,
we conclude that $\Vbend^{(0)}$ has no $\bphi$ dependence
and $\Vbend^{(0)}(\bl)$ satisfies
\begin{equation}
  \dfracp{\Vbend^{(0)}}{\bl}(\bl_{\ast}) = \bzero.
\end{equation}
We can put $\Vbend^{(0)}$ as a part of the spring potential $\Vspring(\bl)$
and neglect it. The included $\Vbend^{(0)}$ modifies the matrix $\mK$
in $O(\epsilon^{0})$.

The first order force is
\begin{equation}
  \left( \dfracp{\Vbend}{y_{\alpha}}(\by) \right)^{(1)}
  = \dfracp{\Vbend^{(1)}}{y_{\alpha}}(\by^{(0)}).
\end{equation}
This force is constant in the fast timescale $t_{0}$.
The first order equations of motion are
\begin{equation}
  \begin{split}
    &
    \dfracp{{}^{2}}{t_{0}^{2}}
    \begin{pmatrix}
      \bl^{(1)} \\ \bphi^{(1)}
    \end{pmatrix}
    +
    \begin{pmatrix}
      \mB_{ll} & \mB_{l\phi} \\
      \mB_{\phi l} & \mB_{\phi\phi}\\
    \end{pmatrix}^{-1}
    \begin{pmatrix}
      \mK_{l} & \mO \\
      \mO & \mO \\
    \end{pmatrix}
    \begin{pmatrix}
      \bl^{(1)} \\ \bphi^{(1)}
    \end{pmatrix}
    \\ 
    & = - 
    \begin{pmatrix}
      \mB_{ll} & \mB_{l\phi} \\
      \mB_{\phi l} & \mB_{\phi\phi}\\
    \end{pmatrix}^{-1}
    \begin{pmatrix}
      \partial_{\bl} \Vbend^{(1)}(\by^{(0)}) \\
      \partial_{\bphi} \Vbend^{(1)}(\by^{(0)}) \\
    \end{pmatrix}.
  \end{split}
\end{equation}
Focusing on the second line of the second term in the left-hand side,
we find no restoring force in $\bphi^{(1)}$.
Therefore, if the gradient of $\Vbend^{(1)}$ at $\by^{(0)}$ is not zero,
secular terms are yielded and they break the perturbation expansion
\eqref{eq:expansion-l-phi}.
Therefore, $\Vbend^{(1)}$ depends on $\bl$ only, and satisfies
\begin{equation}
  \dfracp{\Vbend^{(1)}}{\bl}(\bl_{\ast}) = \bzero.
\end{equation}
As discussed in $O(\epsilon^{0})$, $\Vbend^{(1)}(\bl)$ is also put
in the spring potential $\Vspring(\bl)$
and modifies $\mK$ in $O(\epsilon)$.

Consequently, the leading order term depending on $\bphi$ is $\Vbend^{(2)}(\by)$.
The slow motion of $\bphi$ is hence determined by the second-order bending potential,
which is the same order as the dynamical effects associated
with the averaged term $\mathcal{A}_{\alpha}$.

\section{Simplifications}
\label{sec:simplifications}

We can simplify expressions of the term $\mathcal{A}_{\alpha}$ and the spring energy,
which help to analyze stability of a stationary state.
The idea is to decompose a size-$(2N-2)$ matrix into four half-size submatrices.

\subsection{Decomposition of matrices}

We consider the eigenvalue problem
\begin{equation}
  \mX \mP = \mP \mLambda,
\end{equation}
where $\mX=\mB^{-1}\mK$.
The inverse matrix $\mB^{-1}$ is obtained as
\begin{equation}
  \mB^{-1} =
  \begin{pmatrix}
    \wt{\mB}_{ll} & \wt{\mB}_{l\phi} \\
    \wt{\mB}_{\phi l} & \wt{\mB}_{\phi\phi} \\
  \end{pmatrix}
  \in\MatN,
\end{equation}
where
\begin{equation}
  \begin{split}
    & \wt{\mB}_{ll} = \left( \mA_{C} + \mA_{S}\mA_{C}^{-1}\mA_{S} \right)^{-1}, \\
    & \wt{\mB}_{l\phi} = -\mA_{C}^{-1}\mA_{S}\wt{\mB}_{ll} \mL^{-1} \mS^{\rm T}, \\
    & \wt{\mB}_{\phi l} =  \mS \mL^{-1} \mA_{C}^{-1} \mA_{S} \wt{\mB}_{ll}, \\
    & \wt{\mB}_{\phi\phi} = \mS \mL^{-1} \wt{\mB}_{ll} \mL^{-1} \mS^{\rm T}. \\
  \end{split}
\end{equation}
See Appendix \ref{sec:L-in-y} for the definitions of
the matrices $\mL, \mS, \mA_{C}$, and $\mA_{S}$.
Note that $\wt{\mB}_{ll}\neq \mB_{ll}^{-1}$ in general
and that $\mB^{-1}$ is symmetric.

The decomposition of $\mB^{-1}$ gives
\begin{equation}
  \mX = \mB^{-1} \mK = 
  \begin{pmatrix}
    \wt{\mB}_{ll} \mK_{l} & \mO \\
    \wt{\mB}_{\phi l} \mK_{l} & \mO \\
  \end{pmatrix}.
\end{equation}
The diagonal matrix $\mLambda$ and a diagonalizing matrix $\mP$
are also decomposed as
\begin{equation}
  \mLambda =
  \begin{pmatrix}
    \mLambda_{l} & \mO \\
    \mO & \mO \\
  \end{pmatrix}
  \in \DiagN
\end{equation}
and
\begin{equation}
  \label{eq:P-decomposed}
  \mP =
  \begin{pmatrix}
    \mP_{l} & \mO \\
    \mP_{\phi} & \mE \\
  \end{pmatrix}
  \in \MatN,
\end{equation}
where all the submatrices are of size-$(N-1)$ and $\mE$ is the unit matrix.
The submatrix $\mP_{l}$ solves the eigenvalue problem
\begin{equation}
  \label{eq:eigenproblem-Pl}
  \left( \wt{\mB}_{ll} \mK_{l} \right) \mP_{l} = \mP_{l} \mLambda_{l},
\end{equation}
and the submatrix $\mP_{\phi}$ is determined from $\mP_{l}$ as
\begin{equation}
  \mP_{\phi} = \mS \mL^{-1} \mA_{C}^{-1} \mA_{S} \mP_{l}.
\end{equation}
We further decompose the diagonal matrix $\mW$ as
\begin{equation}
  \mW =
  \begin{pmatrix}
    \mW_{l} & \mO \\
    \mO & \mO \\
  \end{pmatrix}
  \in \DiagN.
\end{equation}

\subsection{Simplification of the spring energy}

The averaged spring energy $\ave{E_{\rm spring}}$ is
\begin{equation}
  \ave{E_{\rm spring}}
  = \dfrac{1}{2} \Tr \left[ \mP^{\rm T} \mK \mP \mW^{2} \right]
  = \dfrac{1}{2} \Tr \left[ \mP_{l}^{\rm T} \mK_{l} \mP_{l} \mW_{l}^{2} \right].
\end{equation}
The matrices of the inside trace are reduced from size $2N-2$ to size $N-1$.
This expression is modified to
\begin{equation}
  \ave{E_{\rm spring}}
  = \dfrac{w(t_{1})^{2}}{2} \Tr \left[ \mP_{l}^{\rm T} \mK_{l} \mP_{l} \mN \right].
\end{equation}
in the use of the hypothesis \eqref{eq:Hypothesis-Matrix}.

\subsection{Simplification of the averaged terms}

Substituting the decomposition of the matrices $\mB, \mLambda, \mP$, and $\mW$
into Eq.~\eqref{eq:mathcalA-averaged} and computing straightforwardly, we have
\begin{equation}
  \label{eq:mathcalA-app}
  \mathcal{A}_{\alpha}
  = \dfrac{1}{4} \Tr \left[ \mP_{l}^{\rm T} \dfracp{\wt{\mB}_{ll}^{-1}}{y_{\alpha}}
    \mP_{l} \mLambda_{l} \mW_{l}^{2} \right],
  \quad
  (\alpha=1,\cdots,2N-2).
\end{equation}
In the way we used the relation
\begin{equation}
  \dfracp{(\mA_{C}\mA_{C}^{-1})}{y_{\alpha}} = \mO
  \quad\Longleftrightarrow\quad
  \mA_{C}^{-1} \dfracp{\mA_{C}}{y_{\alpha}} \mA_{C}^{-1} = - \dfracp{\mA_{C}^{-1}}{y_{\alpha}}.
\end{equation}
Similarly, the function $\mathcal{S}_{\alpha}$ is also simplified to
\begin{equation}
  \label{eq:mathcalS-app}
  \mathcal{S}_{\alpha}
  = \dfrac{\Tr \left[ \mP_{l}^{\rm T} \dfracp{\wt{\mB}_{ll}^{-1}}{y_{\alpha}}
      \mP_{l} \mLambda_{l} \mN_{l} \right]}
  {\Tr \left[ \mP_{l}^{\rm T} \mK_{l} \mP_{l} \mN_{l} \right]},
  \quad
  (\alpha=1,\cdots,2N-2)
\end{equation}
where
\begin{equation}
  \mN =
  \begin{pmatrix}
    \mN_{l} & \mO \\
    \mO & \mO \\
  \end{pmatrix},
  \quad
  \mN_{l} = {\rm diag}(\nu_{1},\cdots,\nu_{N-1}).
\end{equation}
The expressions \eqref{eq:mathcalA-app} and \eqref{eq:mathcalS-app} prove
respectively
\begin{equation}
  \mathcal{A}_{1} = \cdots = \mathcal{A}_{N-1} = 0
\end{equation}
and
\begin{equation}
  \mathcal{S}_{1} = \cdots = \mathcal{S}_{N-1} = 0,
\end{equation}
since the matrix $\wt{\mB}_{ll}^{-1}$ does not depend on $\bl$.

\subsection{Further simplifications}
\label{sec:further-simplification}

The average spring energy and the denominator of $\mathcal{S}_{\alpha}$
are further simplified by choosing a special diagonalizing matrix $\mP_{l}$,
where the matrix $\mP_{l}$ satisfies the eigenvalue problem \eqref{eq:eigenproblem-Pl}.
The symmetric matrix $\mK_{l}$ is positive definite,
and hence we can define the real symmetric matrix $\sqrt{\mK_{l}}$.
Introducing the matrix $\mQ_{l}$ as
\begin{equation}
  \mQ_{l} = \sqrt{\mK_{l}} \mP_{l} \in{\rm Mat}(N-1),
\end{equation}
the eigenvalue problem is rewritten to
\begin{equation}
  \label{eq:eigenvalue-problem-Ql}
  \left( \sqrt{\mK_{l}} ~\wt{\mB}_{ll}~ \sqrt{\mK_{l}} \right) \mQ_{l} = \mQ_{l} \mLambda_{l}.
\end{equation}
The matrix $\sqrt{\mK_{l}} ~ \wt{\mB}_{ll} ~ \sqrt{\mK_{l}}$ is real symmetric,
and hence we can choose $\mQ_{l}\in O(N-1)$,
where $O(N-1)$ is the set of the orthogonal matrices of size $N-1$.
Using $\mQ_{l}\in O(N-1)$, we have
\begin{equation}
  \mP_{l}^{\rm T} \mK_{l} \mP_{l} = \mE.
\end{equation}
This equality simplifies the averaged spring energy to
\begin{equation}
  \ave{E_{\rm spring}} = \dfrac{1}{2} \Tr ~ \mW_{l}^{2}
  = \sum_{i=1}^{N-1} \dfrac{w_{i}^{2}}{2}.
\end{equation}
Energy of the mode $i$ is $w_{i}^{2}/2$ accordingly.
The denominator of $\mathcal{S}_{\alpha}$ becomes
\begin{equation}
  \Tr \left[ \mP_{l}^{\rm T} \mK_{l} \mP_{l} \mN_{l} \right]
  = \Tr~\mN_{l}.
\end{equation}

Moreover, we can introduce the normalization
of the constants $\nu_{i}~(i=1,\cdots,N)$ as
\begin{equation}
  \label{eq:normalization-nu}
  \Tr~\mN_{l} = \sum_{i=1}^{N-1} \nu_{i} = 1,
\end{equation}
since the scale of the spring energy is controled by $w(t_{1})$.
Thus, $\ave{E_{\rm spring}}$ is further simplified to
\begin{equation}
  \ave{E_{\rm spring}} = \dfrac{w(t_{1})^{2}}{2}
\end{equation}
under the hypothesis \eqref{eq:Hypothesis},
and the function $\mathcal{S}_{\alpha}$ is to
\begin{equation}
  \mathcal{S}_{\alpha}
  = \Tr \left[ \mP_{l}^{\rm T} \dfracp{\wt{\mB}_{ll}^{-1}}{y_{\alpha}} \mP_{l} \mLambda_{l} \mN_{l} \right],
\end{equation}
when we use $\mQ_{l}\in O(N-1)$ and the normalization \eqref{eq:normalization-nu}.

\section{Stationarity and stability of one-dimensional conformations}
\label{sec:App-Theory-1Dconf}

We first note that
\begin{equation}
  \label{eq:ASACvanish}
  \mA_{S}(\bphi\in\mathcal{C}^{1}) = 
  \dfracp{\mA_{C}}{\phi_{i}}(\bphi\in\mathcal{C}^{1}) = \mO
  \quad
  (i=1,\cdots,N-1),
\end{equation}
because all the elements depend on $\sin\phi_{i,j}$ in $\mA_{S}$
and $\partial \mA_{C}/\partial\phi_{i}$,
and $\sin\phi_{i,j}=0$ for a conformation belonging to $\mathcal{C}^{1}$.
This fact implies that
\begin{equation}
  \dfracp{\wt{\mB}_{ll}^{-1}}{\phi_{i}}(\bphi\in\mathcal{C}^{1})
  = \left[
    \dfracp{\mA_{C}}{\phi_{i}}
    + \dfracp{(\mA_{S} \mA_{C}^{-1} \mA_{S})}{\phi_{i}}
  \right]_{\bphi\in\mathcal{C}^{1}}
  = \mO
\end{equation}
and
\begin{equation}
  \label{eq:TiC1iszero}
  \mathcal{T}_{i}(\bphi\in\mathcal{C}^{1})=0
\end{equation}
for $i=1,\cdots,N-1$.
Thus, we have $\bG(\bphi\in\mathcal{C}^{1})=\bzero$
for $\Vbend^{(2)}\equiv 0$ from Eq.~\eqref{eq:Gi}.

Similarly, the Jacobian matrix $D\bG$ is simplified as
\begin{equation}
  \dfracp{G_{i}}{\phi_{j}}(\bphi\in\mathcal{C}^{1})
  = - \dfrac{E^{(2)}}{2}
  (B_{\phi\phi}^{-1})^{in}
  \dfrac{\Tr\left[ \mP_{l}^{\rm T} \mY_{nj} \mP_{l} \mLambda_{l} \mN_{l} \right]}
  { \Tr \left[ \mP_{l} \mK_{l} \mP_{l} \mN_{l} \right]}
\end{equation}
where
\begin{equation}
  \mY_{ij}
  = \dfracpp{\mA_{C}}{\phi_{i}}{\phi_{j}}
  + \dfracp{\mA_{S}}{\phi_{i}} \mA_{C}^{-1} \dfracp{\mA_{S}}{\phi_{j}}
  + \dfracp{\mA_{S}}{\phi_{j}} \mA_{C}^{-1} \dfracp{\mA_{S}}{\phi_{i}}.
\end{equation}
We remark that each of $\mY_{ij}~(i,j=1,\cdots,N-1)$ is a size-$(N-1)$ matrix.
Further, the matrix $\wt{\mB}_{ll}^{-1}$ is also simplified to
\begin{equation}
  \wt{\mB}_{ll}^{-1}(\bphi\in\mathcal{C}^{1}) = \mA_{C}(\bphi\in\mathcal{C}^{1}),
\end{equation}
and the matrices $\mLambda_{l}$ and $\mQ_{l}=\sqrt{\mK_{l}}\mP_{l}$
are obtained from the eigenvalue problem
\begin{equation}
  \label{eq:eigenvalue-problem-simplified}
  \left( \sqrt{\mK} \mA_{C} \sqrt{\mK} \right) \mQ_{l}
  = \mQ_{l} \mLambda_{l}.
\end{equation}
Note that we can choose $\mQ_{l}$ from $O(N-1)$.

\section{Eigenvalues of the linearized equations}
\label{sec:extended-eigenvalues}

We consider the eigensystem of the matrix
\begin{equation}
  \mJ =
  \begin{pmatrix}
    \mO & \mE \\
    - \mZ & \mO \\
  \end{pmatrix},
\end{equation}
where all the submatrices are of size-$n$.
We assume that $\mZ$ is diagonalizable.
Suppose that the $n$ eigenvalues of $\mZ$
are real, nonzero, and denoted by $z_{i}$.
The associated $n$ eigenvectors are $\bv_{i}$ satisfying
\begin{equation}
  \mZ \bv_{i} = z_{i} \bv_{i},
  \quad
  (i=1,\cdots,n).
\end{equation}
It is straightforward to show that the matrix $\mJ$
has the eigenvalues $\pm\sqrt{-z_{i}}$ and the associated eigenvectors
$\bv_{i}^{\pm}$ defined by
\begin{equation}
  \bv_{i}^{\pm} =
  \begin{pmatrix}
    \bv_{i} \\ \pm \sqrt{-z_{i}} \bv_{i} \\
  \end{pmatrix}.
\end{equation}

\section{Dynamical stability of one-dimensional conformations by mixed modes}
\label{sec:Theory-mixed-uni}

\begin{figure}[t]
  \centering
  \includegraphics[width=8cm]{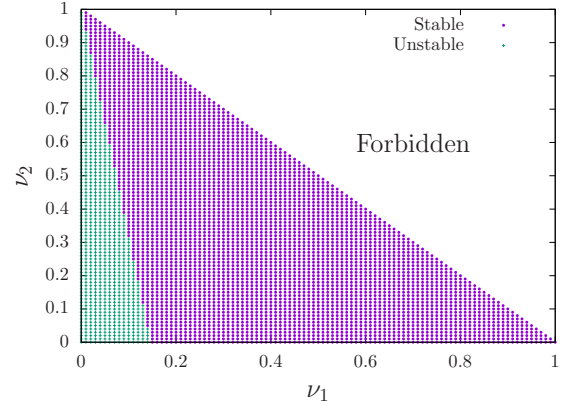}
  \caption{Stable (purple circles) and unstable (green crosses) regions
    on the parameter plane $(\nu_{1},\nu_{2})$ for $N=4$.
    All the one-dimensional conformations share this diagram.
    The parameter $\nu_{3}$ is determined by $\nu_{3}=1-\nu_{1}-\nu_{2}$.
    The critical point on the line $\nu_{2}=0$ is $\nu_{1{\rm c}}\simeq 0.1464466$.
  }
  \label{fig:StabilityN4}
\end{figure}

We study stability of a one-dimensional conformation with exciting multiple modes
under the condition of the equal masses and the identical springs
expressed in Eq.~\eqref{eq:uniform-setting}.
The normal mode energy ratios $\nu_{i}~(i=1,\cdots,N-1)$ are set as
\begin{equation}
  \label{eq:nu-normalization}
  \Tr~\mN = \sum_{i=1}^{N-1} \nu_{i} = 1,
  \qquad
  0 \leq \nu_{i} \leq 1,
\end{equation}
and the number of the free parameters is $N-2$.
We compute $N$ dependence of the stable probability $p_{\rm s}(N)$
with which a considering one-dimensional conformation is stabilized.

A necessary and sufficient condition of the stability for $N=3$ is
\begin{equation}
  \label{eq:ps3}
  \text{the conformation is stable}
  \quad\Longleftrightarrow\quad
  1/4 < \nu_{1} \leq 1
\end{equation}
for the conformations $\bc=(1)$ (straight) and $\bc=(-1)$ (bent).
The condition implies that the probability is
\begin{equation}
  p_{\rm s}(3) = 0.75.
\end{equation}
This stable probability for the two conformations is not a contradiction,
because multi-stability of the two conformations is realized in the interval
\begin{equation}
  1/4 < \nu_{1} < 3/4.
\end{equation}
The condition of Eq.~\eqref{eq:ps3} is in agreement with
the conclusion reported previously \cite{yamaguchi-etal-22}.
The agreement suggests that stability can be obtained by the current theory,
although it does not reduce the rotational symmetry
while the previous theory does.

For $N=4$, we performed numerical computations of stability
at the lattice points $(n_{1}/100,n_{2}/100)~(n_{1},n_{2}=0,\cdots,100)$
on the parameter plane $(\nu_{1},\nu_{2})$,
where $\nu_{3}$ is determined from Eq.~\eqref{eq:nu-normalization}.
The stable and the unstable regions are reported in Fig.~\ref{fig:StabilityN4},
and we have the straight line boundary.
The critical value $\nu_{1{\rm c}}$ on the line $\nu_{2}=0$
is in the interval $[0.1464466,0.1464467]$,
and the stable probability is
\begin{equation}
  p_{\rm s}(4)\simeq 0.8535.
\end{equation}

The stability check on the lattice points is also performed
on the parameter space $(\nu_{1},\nu_{2},\nu_{3})$ for $N=5$.
Among all the researched points of $176851$,
the six conformations are stable at $141019$ points.
Thus, the stable probability is
\begin{equation}
  p_{\rm s}(5) \simeq 0.7974.
\end{equation}
The probabilities $p_{\rm s}(3),p_{\rm s}(4)$, and $p_{\rm s}(5)$ suggest
that dynamical stability is important
even if the system size is large and multiple modes are excited.

\section{Effective potential for $N=3$}
\label{sec:effective-potential}

For $N=3$, the number of the bending angles is two,
and the second bending angle $\phi_{2}$ is not essential,
since it is a cyclic coordinate associating with the total angular momentum.
Reducing $\phi_{2}$, we can construct the effective potential,
which describes the slow bending motion of $\phi_{1}$.
For simplicity, we denote $\phi_{1}$ by $\phi$.
The construction is performed in three steps.

First, instead of $(\nu_{1},\nu_{2})$ defined
in the ascending order of the eigenfrequencies (see the main text),
we use $(\nu_{\rm in}, \nu_{\rm anti})$,
where $\nu_{\rm in}$ ($\nu_{\rm anti}$) represents
the in-phase (antiphase) mode energy ratio.
This change of $\nu$ helps to construct the global effective potential.

\begin{figure}[h]
  \centering
  \includegraphics[width=7cm]{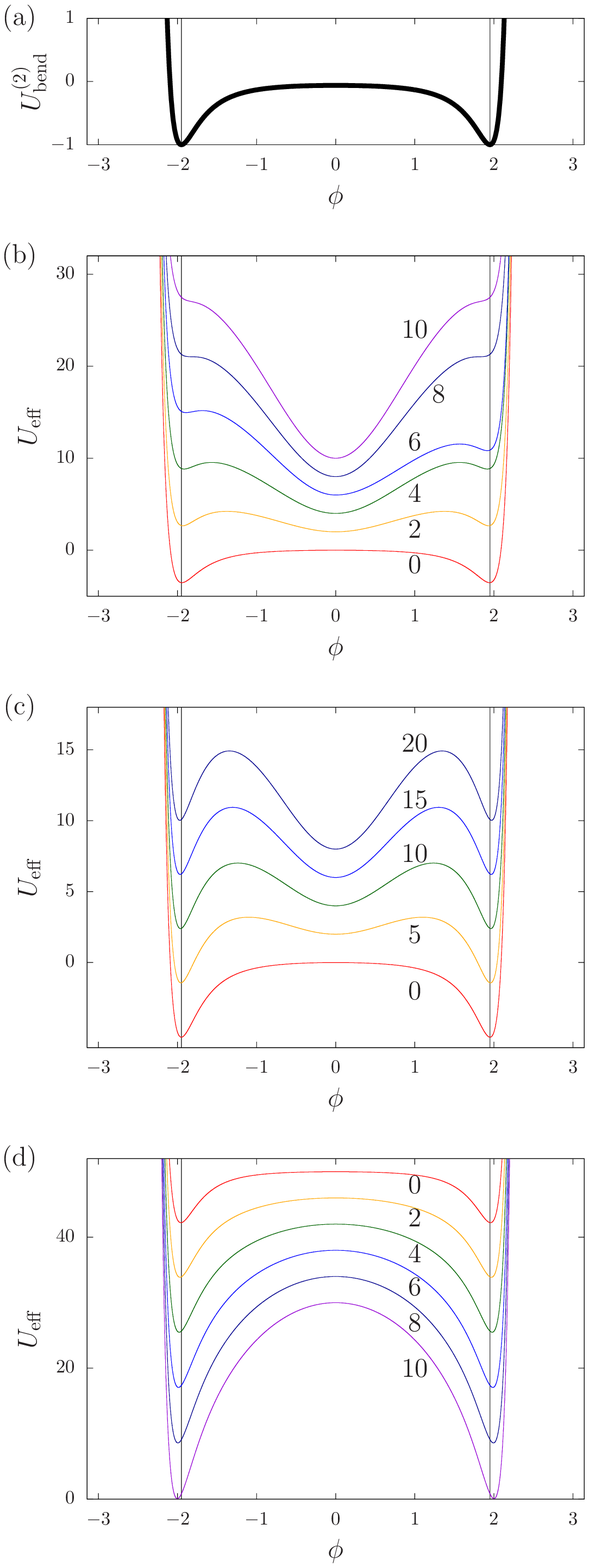}
  \caption{(a) The Lennard-Jones potential $U_{\rm LJ}^{(2)}$
    as a function of the bending angle $\phi$.
    (b) The effective potential $U_{\rm eff}(\phi)$
    with $(\nu_{\rm in},\nu_{\rm anti})=(1,0)$.
    (c) The effective potential $U_{\rm eff}(\phi)$
    with $(\nu_{\rm in},\nu_{\rm anti})=(1/2,1/2)$.
    (d) The effective potential $U_{\rm eff}(\phi)$
    with $(\nu_{\rm in},\nu_{\rm anti})=(0,1)$.
    The numbers in the panels (b)-(d) represent the values of $E^{(2)}$.
    $N=3$,  $\epsilon_{0}=1$, $\sigma=1$, and $l_{\ast}=1$.
    Graphs are vertically shifted for a graphical reason.
    The two vertical lines in each panel mark the minimum points $\pm\phi_{\rm min}$
    of the Lennard-Jones potential.
  }
  \label{fig:ULJ}
\end{figure}

Second, we consider the bending potential
consisting of the interaction between the first and the third beads only.
We set $\Vbend^{(2)}=U_{\rm LJ}^{(2)}$,
and the second-order Lennard-Jones potential $U_{\rm LJ}^{(2)}$
is given in Eq.~\eqref{eq:ULJ2}.
We set $\epsilon_{0}=1, \sigma=1$, and $l_{\ast}=1$.
The graph of $U_{\rm LJ}^{(2)}$ is reported in Figs.~\eqref{fig:ULJ2} and \ref{fig:ULJ}(a).

Third, we construct the effective potential following the previous result
\cite{yamaguchi-etal-22}.
The bending motion is described by the effective Lagrangian
\begin{equation}
  L_{\rm eff}\left( \phi, \dfracd{\phi}{t_{1}} \right)
  = \dfrac{1}{2} M_{\rm eff}(\phi) \left( \dfracd{\phi}{t_{1}} \right)^{2}
  - U_{\rm eff}(\phi).
\end{equation}
The effective mass $M_{\rm eff}$ is
\begin{equation}
  M_{\rm eff}(\phi) = \exp\left[ 2 \int_{0}^{\phi} F(\phi') d\phi' \right]
\end{equation}
and the effective potential $U_{\rm eff}$ is
\begin{equation}
  U_{\rm eff}(\phi) = \int_{0}^{\phi} M_{\rm eff}(\phi') G(\phi') d\phi'.
\end{equation}
The functions $F$ and $G$ are defined by
\begin{equation}
  F(\phi) = \dfrac{1}{2C(\phi)} \dfracp{C}{\phi}(\phi) + \dfrac{1}{4}\mathcal{T}(\phi),
\end{equation}
and
\begin{equation}
  G(\phi) = \dfrac{1}{C} \left[
    \dfracd{\Vbend^{(2)}}{\phi}(\phi)
    - \dfrac{E^{(2)} - \Vbend^{(2)}(\phi)}{2} \mathcal{T}(\phi)
  \right],
\end{equation}
where
\begin{equation}
  C(\phi) = \dfrac{ml_{\ast}^{2}}{6} ( 2 - \cos \phi),
\end{equation}
and
\begin{equation}
  \mathcal{T}(\phi)
  = \left( - \dfrac{\nu_{\rm in}}{2-\cos\phi} 
  + \dfrac{\nu_{\rm anti}}{2+\cos\phi} \right) \sin\phi.
\end{equation}

We can see that the effective potential depends on
the bending potential $\Vbend^{(2)}$,
the mode energy distribution $(\nu_{\rm in},\nu_{\rm anti})$,
and energy $E^{(2)}$.
Examples of the effective potential are exhibited in Fig.~\ref{fig:ULJ}
for three pairs of $(\nu_{\rm in},\nu_{\rm anti})$ with varying $E^{(2)}$.
Excitation of the in-phase mode stabilizes the straight conformation ($\phi=0$),
and the antiphase mode enhances the stability around
the minimum points $\pm\phi_{\rm min}$ of the Lennard-Jones potential
as $E^{(2)}$ increases.

\section{Initial positions of the beads}
\label{sec:initial-position}

We start from a one-dimensional conformation symbolized by $\bc=(c_{1},\cdots,c_{N-2})$.
An initial position with the bending potential $\Vbend$ is prepared
by replacing the bending angle $\phi=\pi$ with $\phi=\pm\phi_{\rm min}$,
where $\phi_{\rm min}$ is a bottom position of the Lennard-Jones potential
[see Eq.~\eqref{eq:phimin} and Fig.~\ref{fig:ULJ2}],
and by approximately exciting normal modes of the one-dimensional conformation $\bc$.

To describe the initial positions of the beads,
we introduce the direction vectors $\bd_{i}~(i=1,\cdots,N-1)$, where
\begin{equation}
  \bd_{1} = (1,0)^{\rm T}.
\end{equation}
We define $\bd_{i}$ as
\begin{equation}
  \bd_{i} = R(\rho_{i}) \bd_{i-1},
  \quad
  (i=2,\cdots,N-1)
\end{equation}
where $R(\rho)$ is the rotation matrix of the angle $\rho$,
\begin{equation}
  R(\rho) =
  \begin{pmatrix}
    \cos\rho & -\sin\rho \\
    \sin\rho & \cos\rho \\
  \end{pmatrix}.
\end{equation}
The angles $\rho_{i}$ are determined as
\begin{equation}
  \rho_{i} = \left\{
    \begin{array}{ll}
      0 & (c_{i} = 1), \\
      - \phi_{\rm min} \prod_{j=1}^{i} c_{j} & (c_{i}=-1), \\
    \end{array}
  \right.
\end{equation}
which replace the bending angle $\phi=\pi$ with $\phi=\pm\phi_{\rm min}$.
Let us denote the initial length of the $i$th spring by $l_{i,0}$,
which is determined later.
The initial position of the $i$th bead $\br_{i,0}$ is
\begin{equation}
  \br_{1,0} = \bzero,
  \quad
  \br_{2,0} = l_{1,0} \bd_{1},
\end{equation}
and
\begin{equation}
  \br_{i,0} = \br_{i-1,0} + l_{i-1,0} \bd_{i-1},
  \quad
  (i=3,\cdots,N).
\end{equation}

The lengths of the springs are decided to excite normal modes
in a desired manner approximately.
As discussed in Appendix \ref{sec:further-simplification},
the matrix $\mP_{l}$ can be constructed as $\mP_{l}=(\sqrt{\mK_{l}})^{-1} \mQ_{l}$,
where $\mQ_{l}\in O(N-1)$ solves the eigenvalue problem \eqref{eq:eigenvalue-problem-Ql}.
For simplicity, we use $\wt{\mB}_{ll}$ defined for the one-dimensional conformation
$\bc$.
Since the $j$th column vector of $\mP_{l}$ represents
the $j$th normal mode of the springs of the one-dimensional conformation $\bc$,
we create the vector
\begin{equation}
  \hat{\bp} = \mP_{l}
  \begin{pmatrix}
    \sqrt{\nu_{1}} \\ \vdots \\ \sqrt{\nu_{N-1}}
  \end{pmatrix}.
\end{equation}
The lengths of the springs are determined as
\begin{equation}
  \bl = \bl_{\ast} + \epsilon \hat{\bp}.
\end{equation}

\end{document}